\documentclass[tradiabstract,longauth]{aa}  
\usepackage{amsmath}
\usepackage[varg]{txfonts}
\usepackage{graphicx}
\usepackage{lscape}
\usepackage{url}
\usepackage{hyperref}

\usepackage{natbib}
\usepackage{multirow}
\usepackage{color}
\usepackage{amssymb}
\usepackage[dvipsnames]{xcolor}
\usepackage{enumerate}
\usepackage{soul} 

\pdfoutput=1

\newcommand{\jp}{J-PAS}
\newcommand{\js}{J-spectra}
\newcommand{\mjp}{miniJPAS}

\def\jphotoz{\texttt{JPHOTOZ}}

\def\magauto{\texttt{MAG\_AUTO}}

\def\totalstar{\texttt{total\_prob\_star}}

\def\zml{\texttt{lephare\_z\_ml}}

\def\baysea{\texttt{BaySeAGal}} 
\def\muff{\texttt{MUFFIT}} 
\def\alstar{\texttt{AlStar}} 
\def\tgas{\texttt{TGASPEX}}

\def\lephare{{\sc LePhare}}
\newcommand{\cb}{C\&B\xspace}

\newcommand{\photoz}{photo-$z$}
\def\rbr{r_\mathrm{SDSS}}
\def\rb{$r_\mathrm{SDSS}$}

\newcommand\ageMt{\langle \log\,\mathrm{age}\rangle_\mathrm{M}}

\newcommand\logZMt{\langle \log\,Z_\star\rangle_\mathrm{M}}

\newcommand\ageM{$\ageMt$}

\newcommand\logZM{$\logZMt$}
\newcommand\ur{$(u-r)_\mathrm{res}$}
\newcommand\urint{$(u-r)_\mathrm{int}$}
\newcommand\Av{$A_V$}
\newcommand\tc{$t_0$}
\newcommand\ta{$\tau$}
\newcommand\tato{\ta/\tc}
\newcommand\sr{\log \Sigma_5}
\newcommand\sur{\log \Sigma_5~[\mathrm{Mpc}^{-2}]}
\newcommand\s{$\sr$}
\newcommand\su{$\sur$}
\newcommand\Par{P_\mathrm{assoc}}
\newcommand\Pa{$\Par$}

\newcommand\logMt{\log M_\star}
\newcommand\logM{$\logMt$}
\newcommand\M{$M_\star$}

\bibpunct{(}{)}{;}{a}{}{,}
\graphicspath{{Figures/}}

\begin{document}

\title{The \mjp \ survey:}
\subtitle{The role of group environment in quenching the star formation}

\authorrunning{Gonz\'alez Delgado et al.}
\titlerunning{Group environment}

\author{
R.~M.~Gonz\'alez Delgado\inst{\ref{IAA}}
\and
J.~E.~Rodr\'iguez-Mart\'in\inst{\ref{IAA}}
\and
L.~A.~D\'iaz-Garc\'ia\inst{\ref{IAA}}
\and
A.~de Amorim\inst{\ref{UFSC}}
\and
R.~Garc\'ia-Benito\inst{\ref{IAA}}
\and
G.~Mart\'inez-Solaeche\inst{\ref{IAA}}
\and
P.~A.~A.~Lopes\inst{\ref{OV}}
\and
M.~Maturi\inst{\ref{ZFA}, \ref{ITP}}
\and
E.~P\'erez\inst{\ref{IAA}}
\and
R.~Cid Fernandes\inst{\ref{UFSC}}
\and
A.~Cortesi\inst{\ref{OV}}
\and
A.~Finoguenov\inst{\ref{DPUH}}
\and
E.~R.~Carrasco\inst{\ref{GEMINI}}
\and
A.~Hern\'an-Caballero\inst{\ref{CEFCA}}
\and
L.~R.~Abramo\inst{\ref{USP-IF}}
\and
J.~Alcaniz\inst{\ref{ON},\ref{UFRGS}}
\and
N.~Ben\'itez\inst{\ref{IAA}}
\and
S.~Bonoli\inst{\ref{CEFCA},\ref{DIPC},\ref{BFS}}
\and
A.~J.~Cenarro\inst{\ref{CEFCA}, \ref{CEFCA2}}
\and
D.~Crist\'obal-Hornillos\inst{\ref{CEFCA}}
\and
J.~M. Diego\inst{\ref{IFC}}
\and
R.~A.~Dupke\inst{\ref{ON},\ref{DAUM},\ref{DPA}}
\and
A.~Ederoclite\inst{\ref{USP}}
\and
J.~A.~Fern\'andez-Ontiveros\inst{\ref{CEFCA}}
\and
C.~L\'opez-Sanjuan\inst{\ref{CEFCA}, \ref{CEFCA2}}
\and
A.~Mar\'in-Franch\inst{\ref{CEFCA}, \ref{CEFCA2}}
\and
I.~M\'arquez\inst{\ref{IAA}}
\and
C.~Mendes de Oliveira\inst{\ref{USP}}
\and
M.~Moles\inst{\ref{CEFCA}, \ref{IAA}}
\and
I.~Pintos\inst{\ref{CEFCA2}}
\and
L.~Sodr\'e Jr.\inst{\ref{USP}}
\and
K.~Taylor\inst{\ref{USP}}
\and
J.~Varela\inst{\ref{CEFCA}, \ref{CEFCA2}}
\and
H.~V\'azquez Rami\'o\inst{\ref{CEFCA}, \ref{CEFCA2}}
\and
J.~M.~V\'ilchez\inst{\ref{IAA}}
}



\institute{
Instituto de Astrof\'{\i}sica de Andaluc\'{\i}a (CSIC), P.O.~Box 3004, 18080 Granada, Spain\newline \email{rosa@iaa.es}\label{IAA}
\and 
Departamento de F\'{\i}sica, Universidade Federal de Santa Catarina, P.O.~Box 476, 88040-900, Florian\'opolis, SC, Brazil\label{UFSC}
\and
Zentrum für Astronomie, Universität Heidelberg, Philosophenweg 12, D-69120 Heidelberg, Germany\label{ZFA}
\and
Institut für Theoretische Physik, Universität Heidelberg, Philosophenweg 16, D-69120 Heidelberg, Germany\label{ITP}
\and
Observat\'orio do Valongo, Universidade Federal do Rio de Janeiro, 20080-090, Rio de Janeiro, RJ, Brazil\label{OV}
\and
Department of Physics, University of Helsinki, Gustaf Hällströmin katu 2, FI-00014 Helsinki, Finland\label{DPUH}
\and
Gemini Observatory/NSF’s NOIRLab, Casilla 603, La Serena, Chile\label{GEMINI}
\and
Centro  de  Estudios  de  F\'isica  del  Cosmos  de  Arag\'on  (CEFCA),  Plaza San Juan 1, E-44001, Teruel, Spain\label{CEFCA}
\and
Instituto de F\'isica, Universidade de S\~ao Paulo, Rua do Mat\~ao 1371, CEP 05508-090, S\~ao Paulo, Brazil\label{USP-IF}
\and
Departamento de Astronomia, Instituto de F\'isica, Universidade Federal do Rio Grande do Sul (UFRGS), Av.~Bento Gonçalves 9500, Porto Alegre, R.S, Brazil\label{UFRGS}
\and
Observat\'orio Nacional, Minist\'erio da Ciencia, Tecnologia, Inovaç\~ao e Comunicaç\~oes, Rua General Jos\'e Cristino, 77, S\~ao Crist\'ov\~ao, 20921-400, Rio de Janeiro, Brazil\label{ON}
\and
Department of Astronomy, University of Michigan, 311 West Hall, 1085 South University Ave., Ann Arbor, USA\label{DAUM}
\and
Department of Physics and Astronomy, University of Alabama, Box 870324, Tuscaloosa, AL, USA\label{DPA}
\and
Donostia International Physics Center (DIPC), Manuel Lardizabal Ibilbidea 4, San Sebasti\'an, Spain\label{DIPC}
\and
Ikerbasque, Basque Foundation for Science, E-48013 Bilbao, Spain\label{BFS}
\and
Instituto de F\'isica, Universidade Federal da Bahia, 40210-340, Salvador, BA, Brazil\label{UFB}
\and
Instituto de F\'isica de Cantabria (CSIC-UC). Avda. Los Castros s/n. 39005, Santander, Spain\label{IFC}
\and
Universidade de S\~{a}o Paulo, Instituto de Astronomia, Geof\'isica e Ci\^encias Atmosf\'ericas, R. do Mat\~{a}o 1226, 05508-090, S\~{a}o Paulo, Brazil\label{USP}
\and
Centro  de  Estudios  de  F\'isica  del  Cosmos  de  Arag\'on  (CEFCA), Unidad Asociada al CSIC, Plaza San Juan 1, E-44001, Teruel, Spain\label{CEFCA2}
}
\date{\today}

\abstract{The \mjp{} survey has observed $\sim 1$~deg$^2$ on the AEGIS field with $60$ bands (spectral resolution of $R \sim 60$) in order to demonstrate the scientific potential of the Javalambre-Physics of the Accelerating Universe Astrophysical Survey (\jp{}) that will map $\sim 8000$~deg$^2$ of the northern sky during the upcoming years. 
In particular, this paper shows the potential of J-PAS to detect groups with mass up to $10^{13}$~$M_\odot$ and the characterisation of their galaxy populations up to $z \sim 1$.
The parametric code BaySeAGal is used to derive the stellar population properties by fitting the J-PAS spectral energy distribution (SED) of the galaxy members in 80 groups at $z \leq 0.8$  previously detected by the AMICO code, as well as for a galaxy field sample retrieved from the whole \mjp{} down to $r<22.75$~(AB).  
Blue, red, quiescent, and transition (blue quiescent or green valley) galaxy populations are identified
through their rest-frame (extinction corrected) \urint{} colour, galaxy stellar mass ($M_\star$), and specific star formation rate (sSFR). We measure the abundance of these galaxies as a function of $M_\star$ and environment to investigate the role that groups play in quenching the star formation. We find: (i) The fraction of red and quiescent galaxies in groups increases with $M_\star$ and it is always higher in groups ($28$~\% on average) than in the field (5~\%). (ii) The quenched fraction excess (QFE) in groups shows a strong dependence with $M_\star$, and  increases from a few percent for galaxies with $M_\star < 10^{10}$~$M_{\odot}$, to higher than $60$~\% for galaxies with $M_\star > 3 \times 10 ^{11}$~$M_\odot$. (iii) The abundance excess of transition galaxies in groups shows a modest dependence with $M_\star$, being $5$--$10$~\% for galaxies with $M_\star < 10^{11}$~$M_\odot$. (iv) The fading time scale, defined as the time that galaxies in groups spend in the transition phase, is very short ($<1.5$~Gyr), indicating that the star formation of galaxies in groups declines very rapidly. (v) The evolution of the galaxy quenching rate in groups shows a modest but significant evolution since $z\sim0.8$. The result is compatible with the expected evolution with constant $QFE=0.4$, which has been previously measured for satellites in the nearby Universe, as traced by SDSS. Further, this evolution is consistent with a scenario where the low-mass star-forming galaxies in clusters at $z= 1$--$ 1.4$ are environmentally quenched, as previously reported by other surveys.
}





\maketitle

\section{Introduction}
\label{sec:Introduction}

Today, the bimodal distribution of galaxy populations is well-known. The Sloan digital sky survey (SDSS) has provided abundant evidence that galaxies in the nearby Universe ($z <$0.1) inhabit two specific loci of the colour-magnitude diagram, the red sequence and the blue cloud \citep{blanton2009}. 
Nearby galaxies in the red sequence are generally characterised by a red, old and metal rich stellar population; whereas galaxies in the blue cloud are mainly star-forming galaxies with blue colours, and with young, less metal rich stars dominating the optical light \citep{kauffmann2003a, kauffmann2003b, baldry2004, brinchmann2004, gallazzi2005, mateus2006, asari2007, gonzalez-delgado2014}. This bimodal distribution of the galaxy populations persists at  higher redshift  \citep{bell2004, williams2009, whitaker2010, hernan-caballero2013, diaz-garcia2019a, diaz-garcia2019b, diaz-garcia2019c, gonzalezdelgado21}. However, since $z \sim 1$, a large fraction of the blue galaxies has seen its star formation being truncated and it has evolved toward the red and quiescent galaxy population, although this is mostly seen for galaxies with  stellar mass lower than $10^{10}$~$M_\odot$, \citep{diaz-garcia2019b}. The transition from blue to red galaxies must occur in a short period of time because the number density of the red galaxies has roughly doubled  since $z\sim 1$ and the number density of the `green valley' galaxies, which lie in between the red sequence and blue clouds objects in the colour-magnitude diagram, is not enough to explain the evolution in number \citep{bell2004, faber2007, muzzin2013}. This  process of transformation from the blue cloud to the red sequence is named `quenching'. 

There is a correlation between the colour and the mass of galaxies belonging to each population. In the local Universe, the galaxies in the red sequence are more massive  than galaxies in the blue cloud \citep{kauffmann2003a, kauffmann2003b, baldry2004, cid-fernandes2005, gonzalez-delgado2014}. However, at intermediate redshifts ($z \sim 1$) this relation is more complex \citep[see e.g. ][]{Ilbert2013}. The stellar mass ($M_\star$) is a relevant galaxy property that correlates with other indicators of the galaxy evolution  \citep{perez-gonzalez2008, perez2013, schawinski2014, lopez-fernandez2018}. In particular, $M_\star$ is related to the quenching process known as mass-quenching \citep{peng2010, kovac2010}. This is based on the strong correlation between the fraction of red galaxies and the stellar mass  \citep{kauffmann2003a, baldry2006}, and the assumption of red colours as a proxy for a quenched stellar  population. Further, galaxies that have shut down their star formation are also referred to as `quenched'. They are out of the relation between the star formation rate and the stellar mass, known as star forming main sequence \citep{noeske2007, speagle2014, renzini2015, gonzalezdelgado2016, lopez-fernandez2018, thorne20}, and have reduced their specific star formation rate (sSFR) by a factor 10--100. Mass quenching is an internal process that is not necessarily caused by the galaxy stellar mass. Other properties, such as the halo-mass \citep[e.g.][]{behroozi2019} or the black-hole mass \citep[e.g.][]{bluck2019, bluck2020}, both of which correlate with \M{}, may be related to mass quenching.

Besides the stellar mass,  the evolution of the galaxy population is also a function of the environment \citep{balogh2004, blanton2005}. Unlike the mass quenching, the environmental quenching is associated with external processes acting in dense environments such as galaxy groups and clusters.  Processes like ram-pressure stripping \citep{abadi1999}, galaxy harassment \citep{moore1996}, starvation \citep{larson1980}, galaxy-cluster tidal interaction \citep{merritt1984, bekki1999}, viscous stripping \citep{nulsen1982}, and thermal evaporation \citep{cowie1977}, can eventually shut down or suppress the star formation by heating and/or removing the gas from the galaxies.  Since the pioneer work by \citet{dressler1980}, many studies have shown the dependence of the distribution of galaxy populations and of their properties with the environmental density \citep[e.g.][]{kauffmann2004, blanton2009, pasquali2010, lopes2014, cappellari2016}.  
There is strong evidence that the fraction of red galaxies increases with the density field for $z< 1$ \citep{woo2013, nantais2016,  darvish2016, calvi2018, moutard2018, liu2021, mcnab2021, sobral2021}, although the quenched fraction is less significant at higher redshift. This result is equivalent to the well-known Bucther-Oemler effect \citep{butcher1984}, which shows that the fraction of blue galaxies in clusters increases with redshift.  

Mass and environmental quenching can be both explained by the halo quenching process, due to the connection between the stellar mass and the environment with the halo mass. Gas in massive galaxy-scale halos ($>10^{12}$ $M_\odot$) is hindered from cooling as it becomes shock-heated \citep{dekel2006}, and the cessation of gas accretion along with ram-pressure stripping prevents further star formation in galaxies, leading the evolution of blue to red galaxies at $z \sim 1$. Moreover, mass and environmental quenching are independent processes because the most massive galaxies are quenched independently of their environment, and galaxies in dense environments are quenched independently of their stellar mass \citep{peng2010}.  Furthermore, \citet{peng2012} and \citet{kovac2014} found little evidence of a dependence of the red fraction of central galaxies on overdensity, whereas satellite galaxies are the main "victims" of environmental quenching in the galaxy population. Satellite galaxies  are consistently redder at all overdensities, and the quenching efficiency
increases with overdensity at $0.1 < z < 0.4$  \citep{kovac2014}. It has been suggested that satellite quenching depends also on the distance to the centre of the halo, and for satellites at lower distance to the halo, quenching depends strongly on the mass halo instead of on the stellar mass \citep{woo2013}.

Galaxy surveys have been very successful at detecting high density structures to study the quenching processes as a function of the global environment (clusters, groups, filaments, voids) or of the local density, through different definitions of the galaxies number density. Good examples are SDSS \citep[e.g.][]{yang2007} and MaNGA \citep{bluck2020} at low-redshift;  and zCOSMOS \citep{peng2012}, CANDELS \citep{liu2021}, GOGREEN \citep{balogh2017, mcnab2021}, CLASH-VLT \citep{rosati2014, mercurio2021} and LEGA-C \citep{sobral2021} at higher redshifts. In general, multi-wavelength photometry and spectroscopic information are combined to get accurate redshifts and to identify the group and cluster galaxy members. Then,  galaxy colours, emission line properties, galaxy mass, and star formation rates (SFR) are derived as proxies to identify the quenched galaxy populations. However, other surveys such as the Hyper Suprime-Cam Subaru Strategic Program (HSC-SSP) combine only deep broad-band photometry with a few narrow band filters to identify groups/cluster galaxy members \citep{lin2017, jian2018, nishizawa2018} and emission line galaxies  \citep[ELG, ][]{koyama2018, hayashi2018}. In this case, H$\alpha$ or [OIII] nebular lines at a given redshift are detected, without distinguishing between recent star formation in the galaxy or  AGN contribution. 

To overcome the limitations of combining broad-band multiwavelength and spectroscopic surveys, the Javalambre-Physics of the Accelerating Universe Astrophysical Survey \citep[\jp, ][]{benitez2009, benitez2014}  was conceived.  \jp{} is a multi-wavelength imaging survey of $\sim 8000$~deg$^2$ in the northern sky that is being carrying out the $2.5$~m telescope
at the Javalambre Astrophysical Observatory (OAJ) \citep{cenarro2014}. 

The photometric system, composes of multi-wavelength narrow bands, is equivalent to a low-resolution spectroscopy of $R \sim 60$. It was designed to measure \photoz\ with an accuracy of up to $\Delta z = 0.003\,(1+z)$ \citep{benitez2014,bonoli21,hernan-caballero21}.  The large area of the survey, the characteristics of the imaging camera, and the accuracy of the \photoz \ allow us to perform multiple cosmological and galaxy evolution studies. It enables to derive the number density of galaxy clusters as a function of redshift and mass to constrain cosmological parameters. 
It is able to map clusters and groups up to $z\sim 1$, and with relatively small masses to produce a complete and mass-sensitive cluster and group catalogue, in order to study the role of environment in galaxy evolution. 

As a low-resolution spectroscopic survey, \jp{} allows us to identify the emission line galaxy population in a continuum range of redshifts up to  $z\sim 1.4$ through [OII], H$\beta$, [OIII], H$\alpha$ and [NII] nebular lines. The emission line diagnostics, such as [NII]/H$\alpha$ and [OIII]/H$\beta$ allow us to discern between AGN and star forming galaxies up to $z\sim 0.35$ \citep{martinez-solaeche2021}. Furthermore, we have already proven the power of \jp{} to identify and characterise the emission-line galaxy population in the AEGIS field \citep{martinez-solaeche2022}  making use of the \mjp{} survey. Briefly, \mjp{} is a proof-of-concept  small
survey covering $\sim$1 deg$^2$,  taken with the same photometric system of \jp{} and the Pathfinder camera \citep{bonoli21}. The \photoz{} are derived for 17500 galaxies per deg$^2$ with $\rbr< 23$, of which $\sim 4200$ have $|\Delta z| < 0.003$ \citep{hernan-caballero21}. Moreover, with \mjp{} we have shown the capability of the \jp{} filter system to dissect the bimodal distribution of red/quenched and blue/star-forming galaxy populations and their evolution up to $z\sim 1$ \citep{gonzalezdelgado21}. 

In this work, we discuss the role of group environment to quench galaxies by comparing the properties of the group members detected in the Extended Growth Strip with the galaxies that are in lower density environments, i.e. in the field. For this we use the \mjp{} survey and the group catalogue constructed by the Adaptive Matched Identifier of Cluster Object (AMICO) \citep{maturi2005}. Our purpose is to show the power of \jp{} to search for the role that the environment plays in the evolution of galaxies; in particular to quench their star formation. The  accuracy of the  \photoz{} \citep{hernan-caballero21} in \mjp{} has allowed us to retrieve a group catalogue  with halo masses down to several times $10^{13}$ $M_\odot$ (Maturi et al. in prep). Further, the analysis of the multi-band data of \mjp{} (\js{}, hereafter) with the full spectra fitting method adapted to multi-narrowband (NB) data, allows us to get the star formation histories of the galaxies to identify the quenched galaxy population in groups and in  lower density environments.

This paper is structured as follows. Section~\ref{sec:Data} briefly describes the data and properties of the sample analysed here. Section~\ref{sec:analysis} explains the method of analysis to fit the \js{}, the classification of galaxies according with the environment, and the identification of the most massive galaxy in each group. In Sect.~\ref{sec:results} we present the inferred stellar population properties of the galaxies, and we compare the properties of galaxies in groups and in the field. Sect.~\ref{sec:fraction-quenched} presents the fraction of red and blue galaxies as a function of the local density and global environment. In Sect.~\ref{sec:Discussion}
we discuss the results in terms of the excess of the quenched fractions in groups, and the evolution of the galaxy quenched rate in groups. Finally, the results are summarised in Sect.~\ref{sec:conclusions}.

Throughout the paper we assume a Lambda cold dark matter ($\Lambda$CDM) cosmology in a flat Universe with $H_0=67.4$~km~s$^{-1}$ and $\Omega_\mathrm{M}=0.315$ \citep{Planck2018}. All the stellar masses in this work are quoted in solar mass units (M$_\odot$) and are scaled according to a universal \citep{chabrier2003} initial stellar mass function. All the magnitudes are in the AB-system \citep{oke1983}.

\section{The data and sample}
\label{sec:Data}

\subsection{Data: observations and calibration}

The observations consist of four pointings in the Extended Groth Strip covering an area of $\sim 1$~deg$^{2}$. The data are from the \mjp\ survey \citep{bonoli21} that were taken with the $2.5$~m Javalambre Survey Telescope (JST/T250) located at the OAJ (Teruel, Spain) \citep{cenarro19} with the Pathfinder camera. The \mjp \ data were obtained with the \jp \ photometric system that contains 54 narrow-band filters of full width at the half maximum (FWHM) of $145$~\AA\ equally spaced every $\sim 100$~\AA\ covering from 3780 to 9100~\AA , plus two mid-band filters centred at 3479~\AA \ and 9316~\AA , and the broad SDSS bands $u, g,r,i$. This system is equivalent to a spectral resolution of $R\sim 60$. 

The observations were processed by the Data Processing and Archiving Unit group at Centro de Estudios de Física del Cosmos de Aragón (CEFCA; \citet{cristobal-hornillos14}) as described in \citet{bonoli21}. 
The \photoz \ were estimated by the \jphotoz{} package developed at CEFCA using a customised version of the \lephare{} code \citep{lephare}. A new set of stellar population synthesis galaxy templates was optimised for the \mjp \ filter system \citep{hernan-caballero21}. The results show that \mjp \, at $\rbr < 23$, has $\sim 17500$ galaxies with valid \jphotoz \ estimates, $\sim 4200$ of which are expected to have $|\Delta z|< 0.003$. All the images and catalogues associated are publicly available at \url{http://archive.cefca.es/catalogues/minijpas-pdr201912}.

\subsection{Sample}
\label{sec:sample}

The galaxy sample used in this work is an extension of that analysed in \citet{gonzalezdelgado21}. It is retrieved from the dual-mode \mjp \ catalogue, selected according to \rb $\leq 22.75$ (\magauto), and redshift ($\mathrm{photo}-z \leq 1$). We also use the `stellar-galaxy locus classification' \totalstar \ parameter \citep{lopez-sanjuan2019, baqui21} listed in the \mjp \ photometry catalogue to select extended sources ($\totalstar \leq 0.5$). We use as \photoz \ the \zml \ parameter listed in the \mjp \ catalogue provided by \citep{hernan-caballero21}, that is the median redshift of the probability density function (PDF) of the \photoz \ distribution for each object. 
In total, we select 11281 objects; and for 99~\% of the galaxies in the sample we were able to get a reasonable SED-fit (see Sect.~\ref{sec:analysis}). These selection criteria are very similar to those used by (Maturi et al. in prep.) to identify galaxy groups and clusters in \mjp.


\section{Analysis}
\label{sec:analysis}

\subsection{The \js \ fits}
\label{sec:method}

\begin{figure*}
\centering
\includegraphics[width=0.8\textwidth]{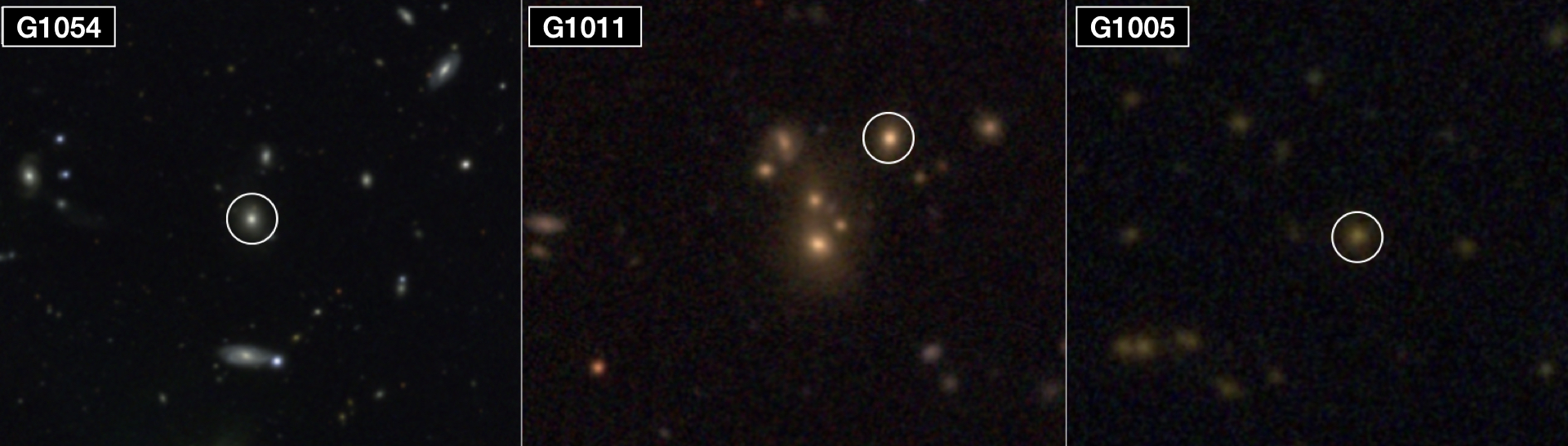}
\includegraphics[width=\textwidth]{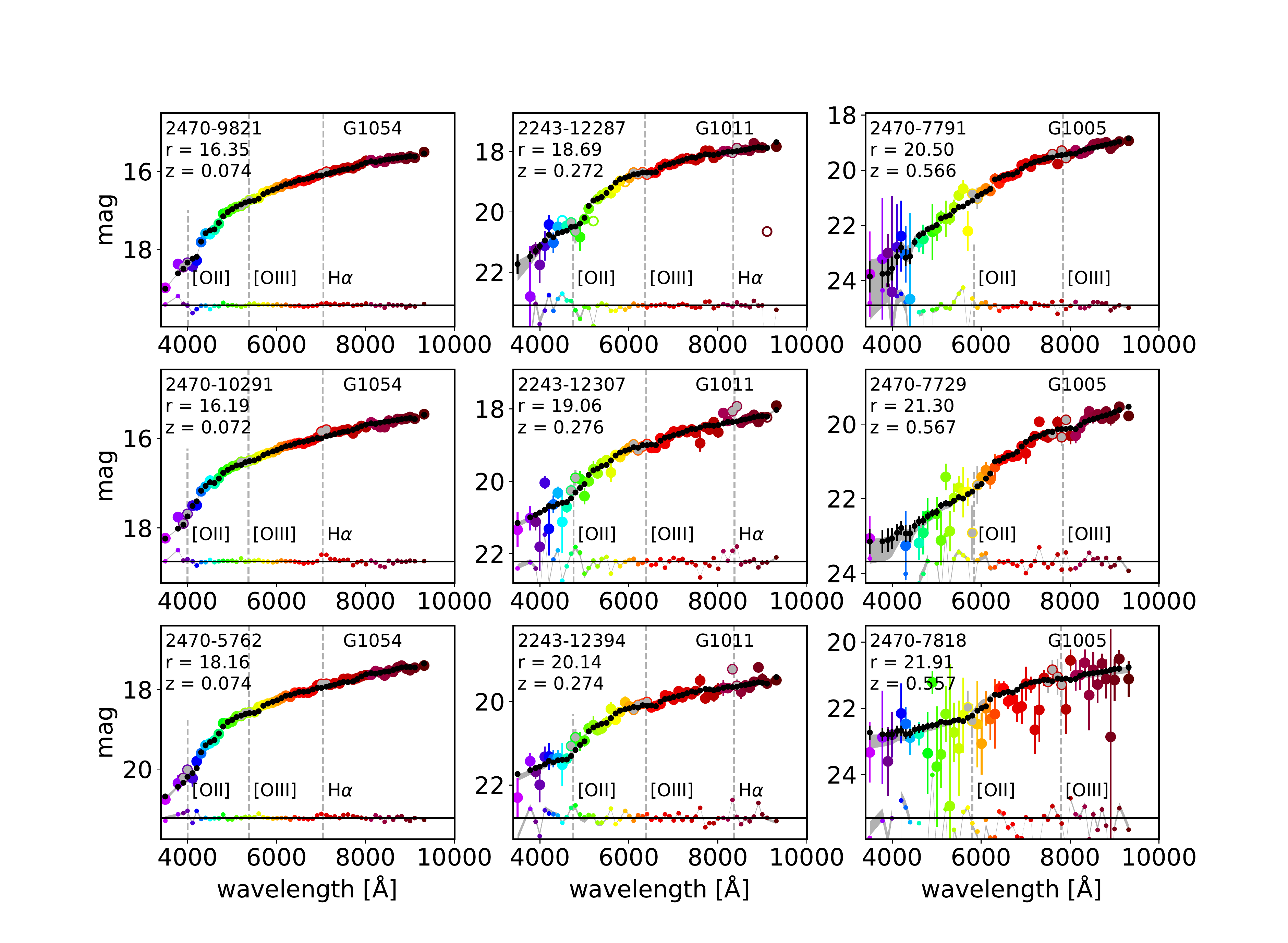}
\caption{Images and \js{} \ of several galaxy members in three AMICO groups. {\it Upper panel:} Images of the central part of G1054, G1011, and G1005. The most massive galaxy in each group is marked by a circle.
{\it Middle and bottom panels:} \js{} (\magauto{}) of galaxies members of G1054, G1011, and G1005, as labelled. The most massive galaxy (marked with a circles in the upper panel), the second most massive galaxy in each group, and other galaxy member of the group are shown.
\js{} are shown as coloured dots, while the best model fitted by \baysea{} for each \js{} is plotted as black points; the grey band shows the magnitudes of the mean model $\pm$ one $\sigma$ uncertainty level.
The differences between the observed and the magnitudes of best model fitted are plotted as a small coloured points around the black bottom line.
Masked filter (white coloured circles) and filters overlapping with the emission lines H$\alpha$, [NII], [OIII], H$\beta$ and [OII] (grey coloured circles) are not used in the fits. The dashed vertical lines shows the wavelength positions where could be H$\alpha$, [OIII], and [OII] in emission at the redshift of each galaxy. The H$\alpha$ line is clearly detected in the galaxies 2243-12307 and 2243-12394 that belong to G1011, and 2470-10291 of G1054. 
}
\label{fig:jspectra}
\end{figure*}

To estimate the stellar population properties of the galaxies as a function of the environmental conditions we fit the \js{} with a SED-fitting code. Fig.~\ref{fig:jspectra} shows several examples of galaxies' \js{} that belong to three AMICO groups,  
at redshift $\sim$ 0.07, 0.27 and 0.57.  Red galaxies and galaxies with H$\alpha$ emission (pressumably star-forming galaxies) are present in these groups. The  S/N ratio and the quality of the \js{} are very similar for the red and blue galaxy populations in \mjp{} \citep{gonzalezdelgado21}.  

Here, we use the SED-fitting code \baysea{} \citep{gonzalezdelgado21} to fit the \js{}. This is a Bayesian parametric code that assumes the latest versions of the \citet{bruzual2003} stellar population synthesis models
\citep[][hereafter \cb]{plat2019}. The \cb models follow the PARSEC evolutionary tracks \citep{marigo2013,chen2015} and use the Miles \citep{sanchez-blazquez2006,falcon-barroso2011,prugniel2011} and IndoUS \citep{valdes2004,sharma2016} stellar libraries in the spectral range covered by the \js{} data.  

The code assumes a  $\mathrm{SFH}\footnote{Star Formation History} = \mathrm{SFH}(t; \Theta)$, where $t$ is the lookback time and $\Theta$ is a parameter vector that includes the stellar metallicity ($Z_\star$), a dust attenuation parameter (\Av{}), and the parameters ($k$, \tc{}, \ta{}) that control the time evolution of the SFR\footnote{Star Formation Rate}, $\psi(t)$. We assume a delayed-$\tau$ model of the form:

\begin{equation}
\psi (t) = k \frac{t_0-t}{\tau}\exp{ \left[ -(t_0-t)/\tau \right] }, 
\end{equation}
where \tc{} is the time of the onset of the star formation in lookback-time, \ta{} is the SFR e-folding time and $k$ is a normalization constant related to the total mass formed in stars. The galaxy stellar mass is calculated from the mass converted into stars according with the SFH and the luminosity of the galaxy, and taking into account the mass loss of the single stellar population synthetic models owing to stellar evolution.

\baysea{} follows a Markov chain Monte Carlo (MCMC) approach, that explores the parameter space and constrains the parameters of the SFH that fit the \js{} . The code allows us to retrieve the chains and their $\chi^2$ likelihood, to
derive the PDF for each of the stellar population properties (stellar mass, stellar age, dust attenuation, stellar metallicity, and colours), and the median and sigma of the chains as its inferred value and error.

The  solutions from the \js{}-fits reproduce the 56 \mjp{} magnitudes of galaxies of different types quite well within the uncertainties and independently of the redshift and brightness range (e.g.~Fig.~\ref{fig:jspectra}). The emission lines from young star-forming regions and/or AGN contributions are not fitted with this code. Therefore, the NBs affected by the contribution of the most relevant lines, H$\alpha$, H$\beta$, [NII]$\lambda$6584, 6548, [OIII]$\lambda$5007, 4959, and [OII]$\lambda$3727, are removed from the \js{} fit at the redshift of each galaxy. Thereby, the fits are restricted to the stellar continuum. 

A more detailed explanation about the inputs and assumptions in the code, and how the results compare with other codes such as \muff{} \citep{diaz-garcia2015}, \alstar{} (Batista et al. in prep.), and \tgas{} \citep{magris2015} can be found in \citet{gonzalezdelgado21}. 


\begin{figure*}
\centering
\includegraphics[width=0.98\textwidth]{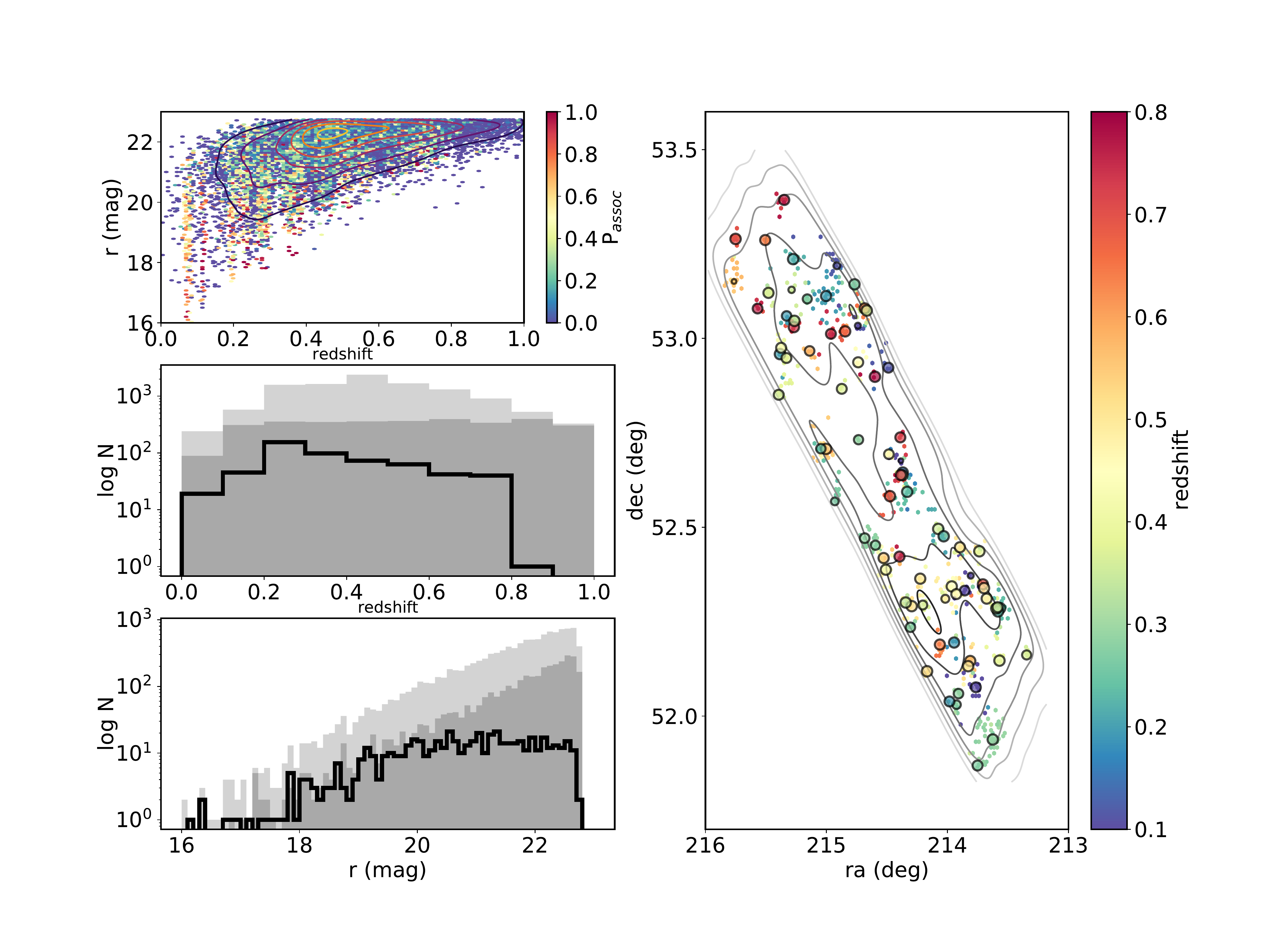}
\caption{Observational properties of the sample and distribution of the AMICO galaxy members. Upper-left panel: redshift--magnitude diagram for the whole sample coloured by the probabilistic association, \Pa{}. Middle-left panel: the redshift distribution. Bottom panel panel: \rb \ (\magauto) distribution. They show the galaxies in group environment (black line), and galaxies in the field (grey histograms). The distributions of the full \mjp \ sample of galaxies are also shown (light grey histograms). 
Right panel: The contours show the density galaxy map distribution of whole \mjp \ sample analysed here. The points 
show the distribution of galaxies in groups; the circles are the brightest and most massive galaxy of each AMICO group. Points are coloured by the \photoz .
}
\label{fig:maps}
\end{figure*}

\subsection{Galaxy classification vs environment}
\label{sec:galaxymembers}

The Adaptive Matched Identifier of Cluster Object (AMICO) code \citep{maturi2005, AMICO} is used in \mjp \ for the detection of galaxy groups and clusters. The code is based on an optimal filtering approach, which minimises the noise variance under the condition that the estimated signal is unbiased. Using as input the redshift of the galaxies, their magnitudes, sky position, and the background noise, the code provides the amplitude  and the association probability (\Pa{}) for each galaxy to be member of a cluster or group. Because the different clusters can overlap in the data space, more than one cluster association can be assigned to the same galaxy through an iterative approach, being the new $\Par =1-\sum_k^{j-1} \Par(k)$, 
where the sum is extended to the probabilities of the previous clusters/groups assignments. This is a key parameter in our study because it allows us to identify the galaxy members of a group or cluster, and hence to characterise the galaxy populations in terms of the global environment.

The good performance of AMICO and \mjp \ regarding mass sensitivity, mass-proxy quality and redshift accuracy show that \jp \ will allow us to derive cosmological constraints not only based on cluster counts, but also on clustering of galaxy clusters (Maturi et al. in prep.).  From this analysis, AMICO  has identified $\sim 80$ groups in \mjp \ at $z < 1$ down to $10^{13}$~$M_\odot$,  when the \photoz{} \ defined as the \zml \ is taken as the galaxy redshift, and the \rb \ (\magauto ) as inputs to the code. 

To identify and characterise the galaxy populations in terms of group environment, we assign a \Pa{} to each galaxy of the sample. For this purpose we perform a cross match of the catalogues with the galaxy group members and the \mjp \ galaxies. As mentioned before, a galaxy can have a \Pa{} larger than 0 in more than one of the galaxy groups in the catalogue, so in our analysis we only consider for this galaxy the highest \Pa{} among those.
If a galaxy is not listed in any of the group catalogues, we set the \Pa{} of this galaxy equal to zero. 

Roughly half of the galaxies of the sample (49~\%) are not listed in any of the group catalogues, indicating that they are not within a group environment; in contrast, only 14~\% of the \mjp{} galaxies have  $\Par \geq 0.5$, and 7~\% have $\Par \geq 0.7$. We use  \Pa{} to segregate the galaxy populations in two different environments: galaxies in groups if  $\Par \geq 0.7$,  and galaxies in the field  if $\Par \leq 0.1$.  They are two sub-samples very different in number, $7$~\% and $63$~\% of the galaxies belong to group and field environment, respectively. However, they show similar range in  magnitude (Fig.~\ref{fig:maps}). In term of redshift, a few galaxies in groups are detected at $z>$ 0.8. As expected, the galaxies in groups trace the most dense areas of \mjp \ galaxy populations (right panel in Fig.~ \ref{fig:maps}). In addition to these two sub-samples, we consider the galaxies with 0.1 $< \Par{} <$ 0.7 as part the whole AEGIS sample.

\subsection{Group stellar mass }
\label{sec:groups}

\begin{figure*}
\centering
\includegraphics[width=0.49\textwidth]{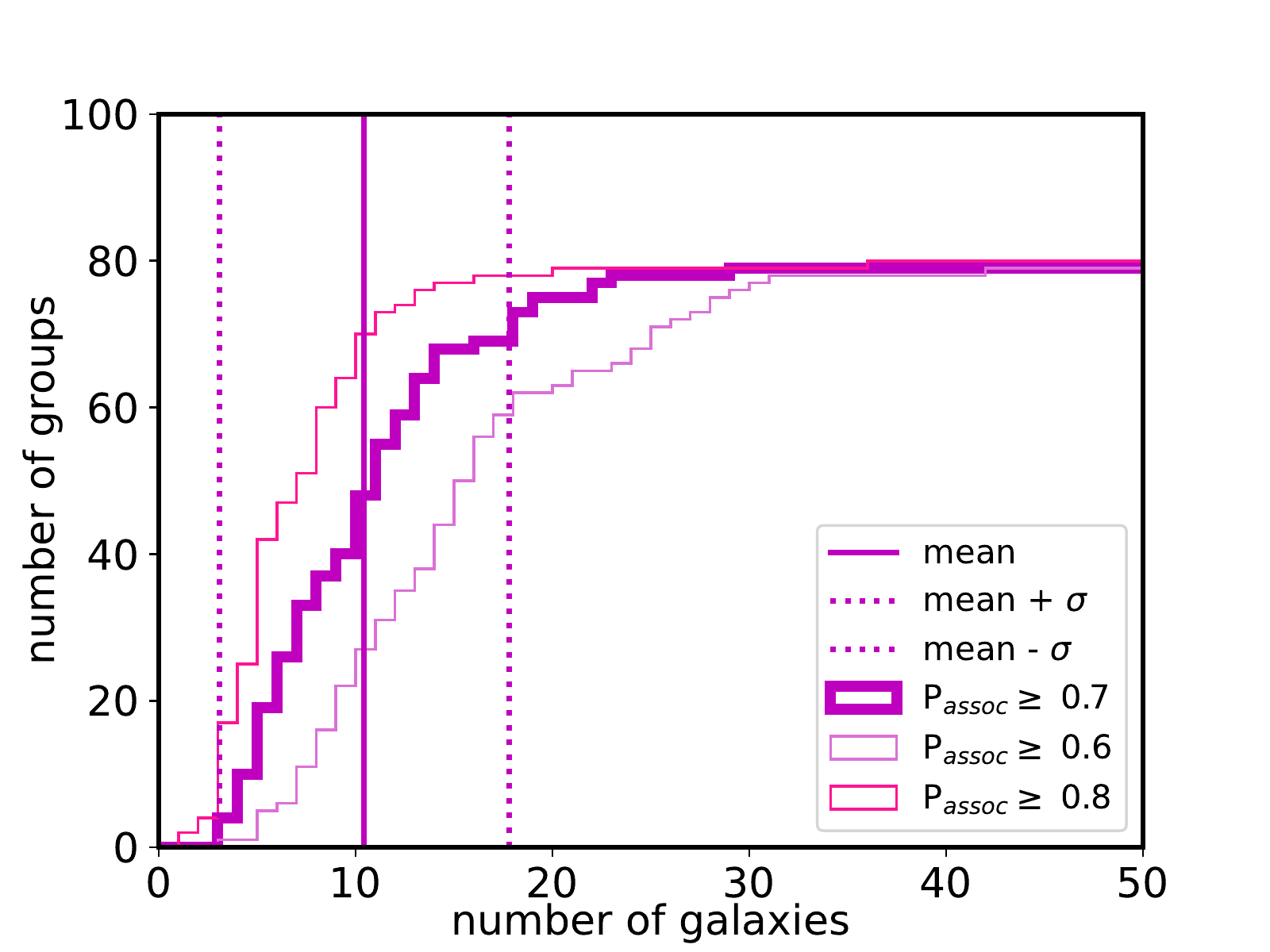}
\includegraphics[width=0.49\textwidth]{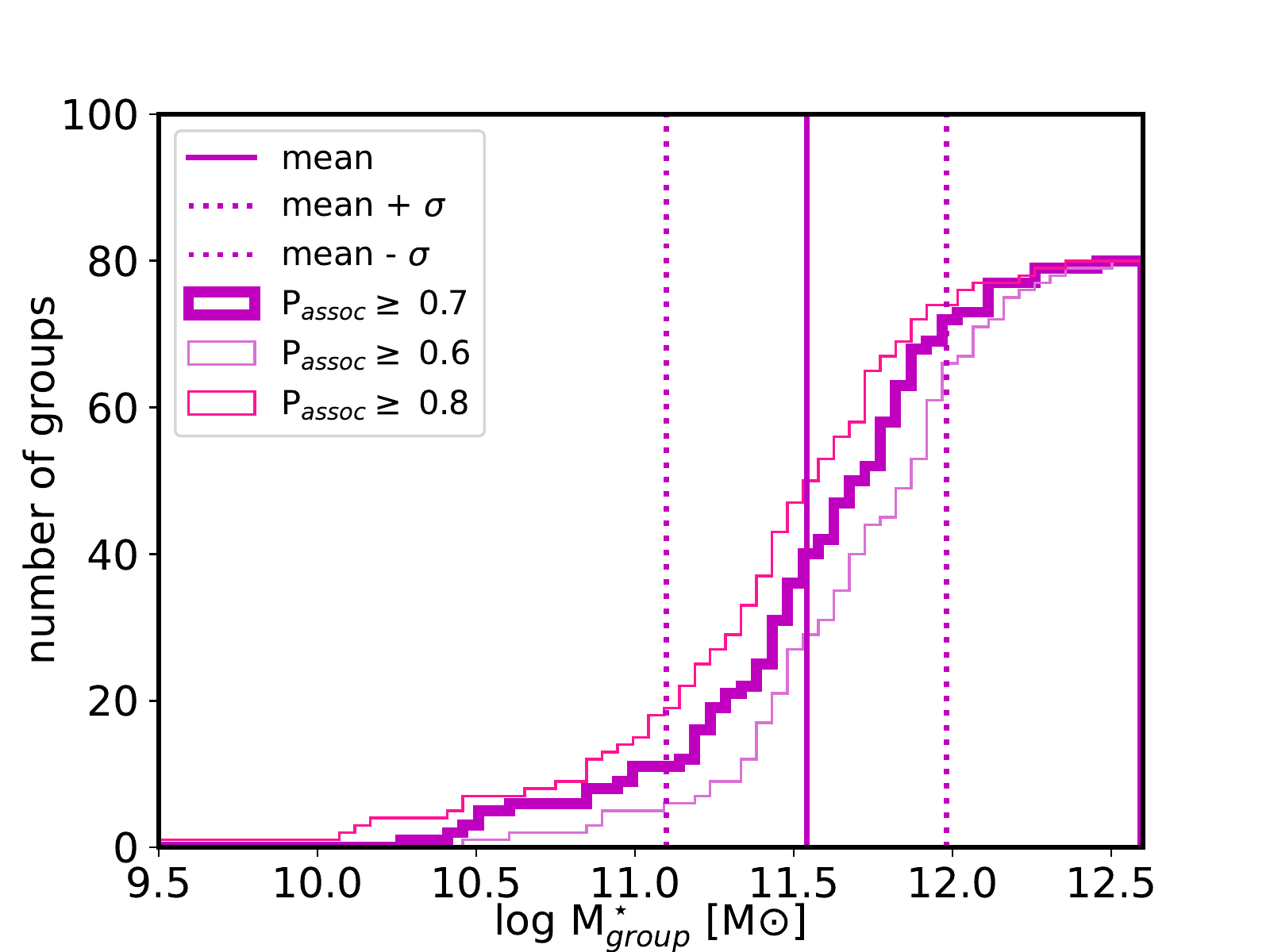}
\caption{Left panel: Cumulative distribution of the number of galaxy members in the AMICO groups. The distribution is derived for three values of \Pa{} as indicated in the inset. The vertical solid line is the average number of galaxy members per group; the dashed lines show $\pm$1 sigma. Right panel: Cumulative distribution of the group stellar mass.   }
\label{fig:galaxymembers}
\end{figure*}

Most of the high density structures detected by AMICO have a number of galaxies that is typical of galaxies groups. As Fig.~\ref{fig:galaxymembers} shows, half of the structures have less than 10 galaxies per group (typical value = $10.4\pm7.4$), which is a number more typical of groups than clusters. This number, however, shows a strong dependence on \Pa{} (Fig.~\ref{fig:galaxymembers}). It varies from $\sim 5$ ($\Par \geq 0.8$ ) to $\sim 15$ ($\Par \geq 0.6$) galaxy members. However, even the highest values are still below the typical value of galaxy members in clusters. One exception, however, is G1001, named mJPC2470-1771 (Rodr\'\i guez-Mar\'\i n et al. 2022, submitted). This is the only one that has $>50$ galaxy members, and it can be considered a cluster (the most massive \mjp{} cluster).

The halo masses obtained through the scaling relation have values in the range of galaxy groups (in the order of $10^{13}$~$M_\odot$) (Maturi et al. in prep.). It is also known that the mass of the
dark matter halo associated with a group is well correlated with  the total stellar mass of the group, and with the mass of the most massive galaxy in the group \citep{yang2007}. 
Here, we can estimate the stellar mass of each group by adding the individual galaxy stellar mass. 
It is worth noting that the stellar mass of a galaxy is expected to be $\sim 1$--$2$~\% of its `halo' mass \citep{moster2010, behroozi2013}.
Figure~\ref{fig:galaxymembers} shows the distribution of the group stellar mass of the AMICO groups. Half of the AMICO groups have log M$^\star_\mathrm{group}$ $\leq 11.5$~[$M_\odot$]. Only, mJPC2470-1771 has log M$^\star_\mathrm{group}$ $> 12$~[$M_\odot$], and a halo mass from the scaling relation which is above $10^{14}$~$M_\odot$. 
In contrast to the number of galaxies in each group, M$^\star_\mathrm{group}$ show a weak dependence with \Pa{} for $0.6\leq \Par \leq 0.8$ (Fig.~\ref{fig:galaxymembers}). 
Thus, we can conclude that, given the number of members and the stellar mass, these AMICO structures are groups. However, M$^\star_\mathrm{group}$ is  $\sim$0.11 dex lower than the total group mass calculated by weighting each galaxy $M_\star$ by its \Pa{} and adding all the galaxies with \Pa{}$>$0.5.

\subsection{Identification of the most massive-brightest-central galaxy in each group}
\label{sec:BGG}

\begin{figure*}
\centering
\includegraphics[width=\textwidth]{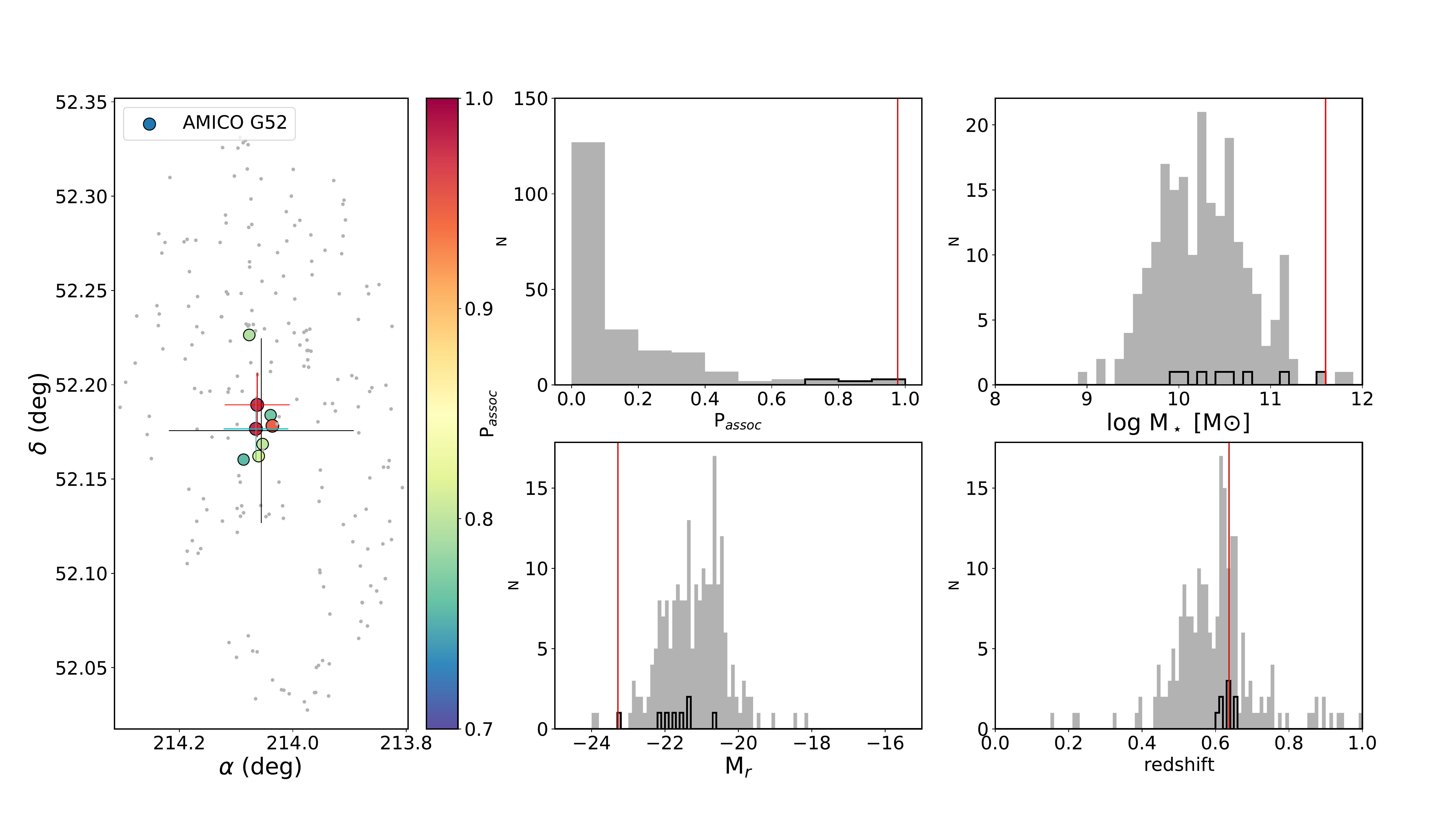}
\caption{Distribution in the sky of the galaxies members listed in the catalogue of the AMICO group G52 (left panel). Objects in the catalogue with \Pa{}$<$0.7 are plotted by grey dots. The galaxies with \Pa{}\ ($\geq$0.7)are shown by circles that are coloured with their \Pa{}. The most massive galaxy, and the galaxy with the highest \Pa{} are marked by a red and a cyan cross, respectively. The group center is also marked by a big black cross. The right panels show the distribution of \Pa{} in the catalogue, the stellar mass, absolute magnitude in the $r$ band, and redshift. The position of the BGG is marked by a red vertical line. The distribution of the galaxy members (\Pa{} $\geq$ 0.7) is shown by the black histogram. }
\label{fig:example}
\end{figure*}

The brightest cluster galaxy (BCG) is the brightest galaxy within the high density structure, that is located at the geometrical and kinematic centre of the cluster if it is in equilibrium. Usually, it is a massive and red early type galaxy at the centre of the potential well, that in many cases is coincident with the maximum of the X-ray emission. However, the structures found in \mjp{} by AMICO are not so big; they are mainly groups, as we have already pointed out.  The identification of the brightest and central galaxy of each group is not easy as the brightest and the most massive galaxy do not have to be at the centre of the structure.

First, we need to determine the geometrical centre of the group. For this we only use the first five galaxies with the highest \Pa{} that are listed in the AMICO catalogue of each group. Then, a distance probability ($P_{\mathrm{dist}}$) is associated to each galaxy group member, that scales with the distance in linear way that this is equal to 1 for the galaxy that is closest to the centre, and zero for the galaxy that is at the largest distance of the group centre. Similarly, we associated to each galaxy a mass probability ($P_\mathrm{mass}$), that scales with the galaxy mass to be equal to 1 for the most massive galaxy of the group, and zero for the less massive galaxy in the group. Once defined these normalised probabilities, we chose as the central/massive/brightest galaxy of each group the galaxy member that has the highest $P = \Par \times P_{\mathrm{dist}} \times  P_\mathrm{mass}$. For many of the groups, this galaxy coincides with the most massive galaxy, and the brightest galaxies of the group (BGG). We note that, in general, there may not be a galaxy in the geometrical centre of the group, and that the BGG may not be the galaxy with the highest \Pa, or 
the brightest galaxy in the group, but its mass is similar to the most massive one. One example is presented in Fig.\ref{fig:example} for the AMICO group G52.  The galaxy  `2470-13620'  is one of the galaxies with highest \Pa, the most massive galaxy, and the brightest galaxy in the group; however, it is not the one closest to the centre.

\subsection{Local density of galaxies in the group environment}
\label{sec:sigma5}

\begin{figure*}
\centering
\includegraphics[width=0.49\textwidth]{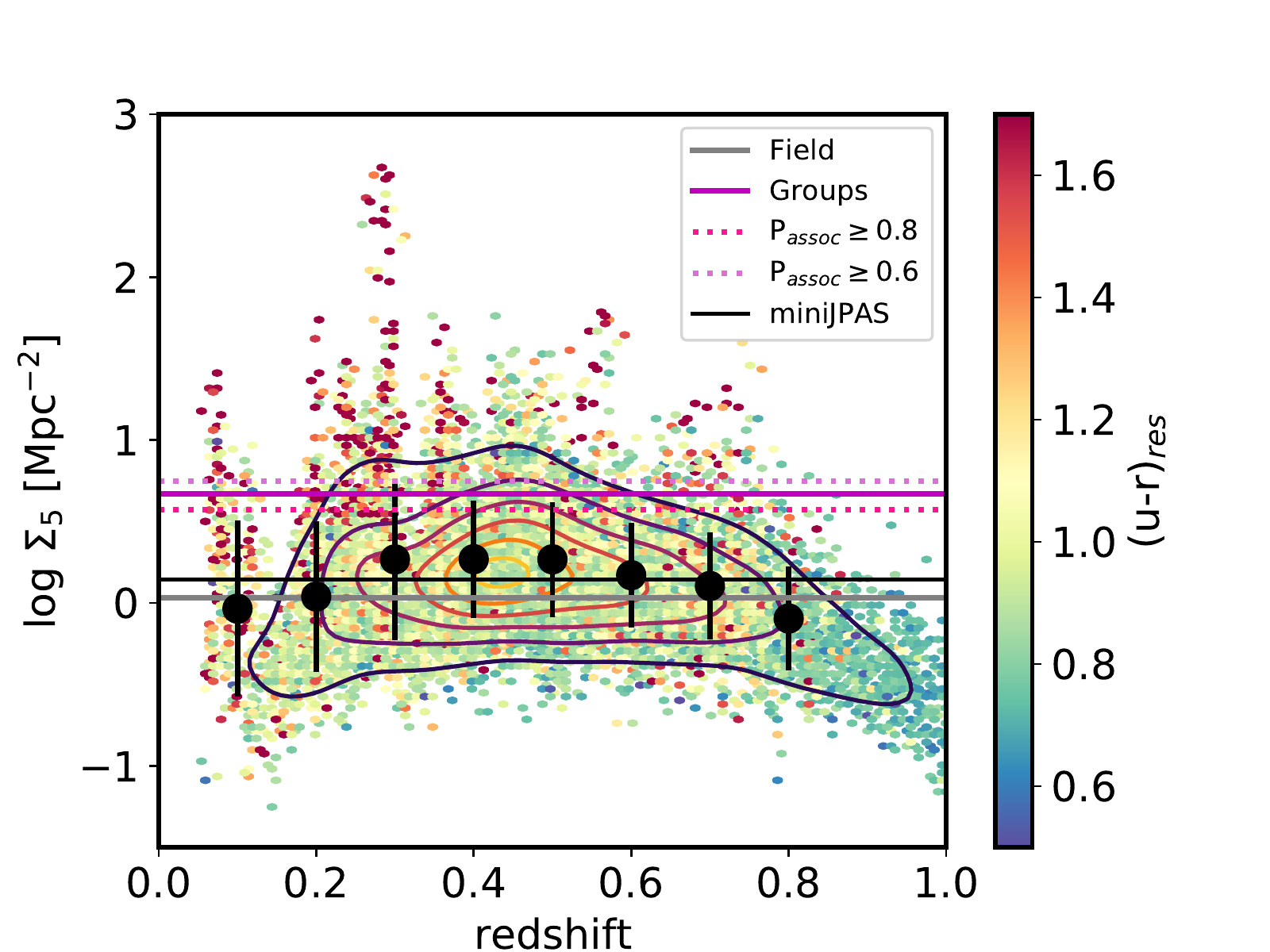}
\includegraphics[width=0.49\textwidth]{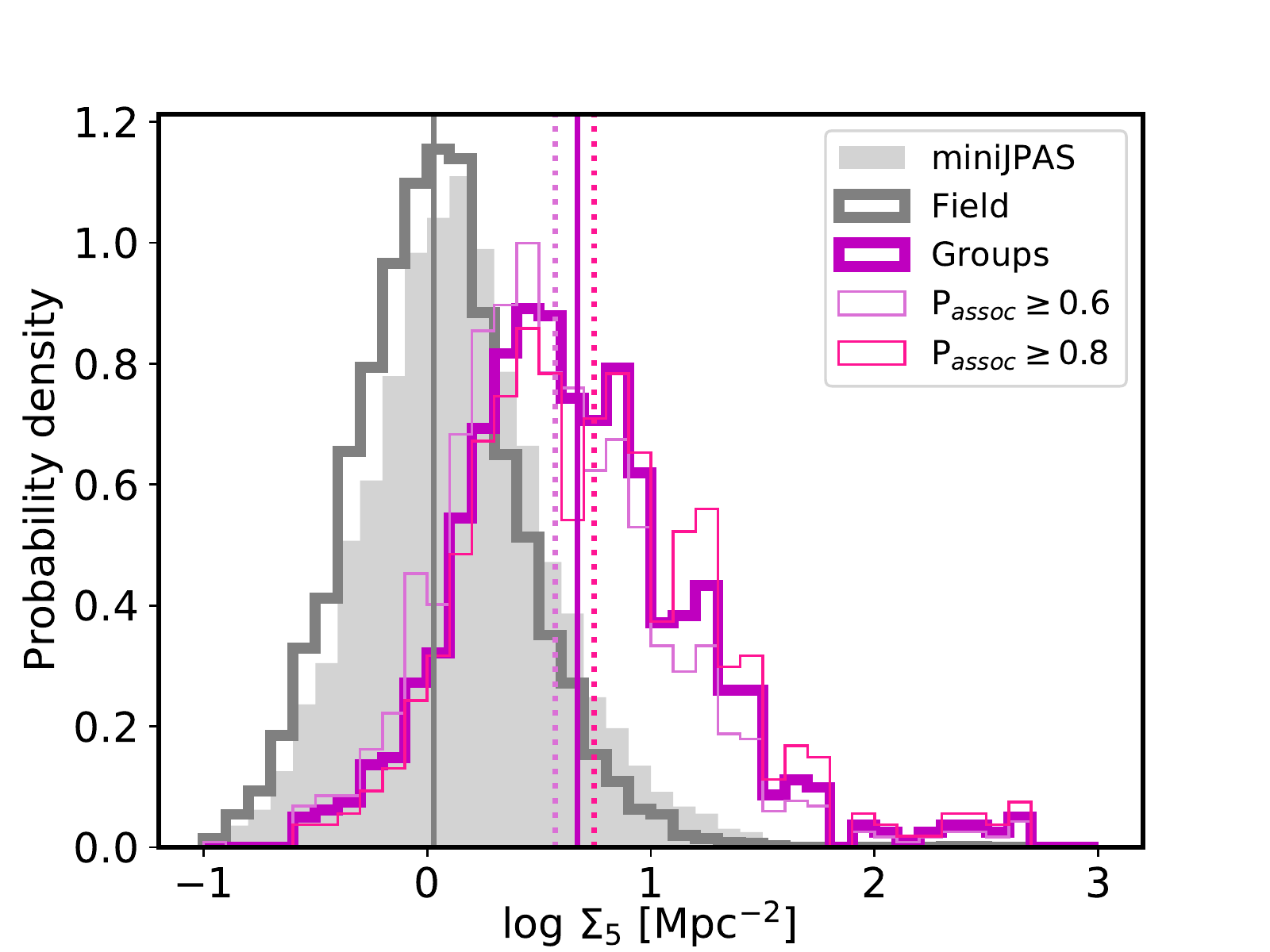}
\caption{{\it Left panel:} The distribution in redshift of \s{}. The contours show the density distribution of points. The average values in different redshift bins are shown as black points. The black line shows the average \s{} for all the \mjp \ galaxies; magenta and blue lines, the average \s{}\ for galaxies in groups and in the field environment, respectively. Dashed lines are at the average \s\ for galaxies with $P_\mathrm{assoc} > 0.6$ or $0.8$. The colour bar shows the distribution of the rest-frame $(u-r)$ colour. 
Vertical structures are groups in \mjp{}.
{\it Right panel:} Distribution of \s{} for \mjp \ galaxies (grey area), and field (grey line) and galaxies in group (magenta) environment. Vertical lines show the positions of the average values. 
}
\label{fig:sigma5}
\end{figure*}

The local density is a proxy of the local environment, which is sensitive to the processes taking place on small scales \citep[e.g.][]{calvi2018}. It can be defined in different ways \citep[e.g.]{Muldrew12}; for example, by counting galaxies within a fixed radius (such as 0.5 Mpc or 1.0 Mpc) or by measuring the distance to the N-th nearest neighbour ($d_N$). Usually, N between  3 to 5 is enough  to characterise  small scale, while  N$\sim$10 is for large scale and denser environment.

To characterise the local density of galaxies in \mjp{}, we use the environment indicator $\Sigma_5$ \citep{lopes2016}.  It is defined by $\Sigma_5 = 5/(\pi d_5^2)$ and describes the local number density around a galaxy within an area defined by the projected area of the 5th nearest neighbor ($d_5$) within a given redshift slice \citep{dressler1980}. With  this definition, the local density is measured in units of galaxies per Mpc$^{-2}$. 

The distribution of \s{}[Mpc$^{-2}$] for the \mjp{} galaxy sample ranges from $-1$ to $2.5$,  with an average value of $0.1$  (Fig.~\ref{fig:sigma5}). This mean value is low, and lower than the local density expected for galaxies located in the center of clusters \su{} $\geq 2$ \citep[see e.g. ][]{Lopes2020}. The fraction of galaxies above \su $\geq 2$  is small ($< 0.2$~\%), suggesting that certainly there are not too many high overdensity structures, such as clusters, in \mjp .  However, the galaxies in AMICO groups are tracing the local overdensity in \mjp , with an average and dispersion values of \su{} equal to $0.67 \pm 0.52$. We stress that the distribution of \s{} for galaxy group members changes very little with \Pa{} (see panel right in Fig.~\ref{fig:sigma5}). The average of \s{} changes from $0.57$ to $0.75$~$\mathrm{Mpc}^{-2}$ when galaxies with \Pa{} larger than $0.6$ or $0.8$ are taken as galaxy members. Thus, the galaxy members of the AMICO groups are tracing intermediate densities, and they are very useful to study the role that the group environmental conditions play in quenching the star formation in galaxies. It is also interesting to note that the BGGs of these groups are tracing a similar distribution of the local overdensity as the  satellites galaxies do, since they are found in the regions with higher contour densities (see right panel of Fig.~\ref{fig:maps}), also with an equal average value of \s{} of 0.64 (std = 0.53). On the other hand, the galaxies in the field have a average  \su{}   of 0.03 (std = 0.39), that is significantly lower than the average density of galaxies in groups. Notice that although our definition of the local environment is correlated with the election of field and group galaxies members, there is some overlap between both populations (right panel in Fig.~\ref{fig:sigma5}). There are field galaxies located in overdensity regions with \su{} $>$0.5 and some galaxy group members in regions of \su{} $<$0.

\section{Stellar population properties of galaxies in groups}
\label{sec:results}


\begin{figure*}
\centering
\includegraphics[width=\textwidth]{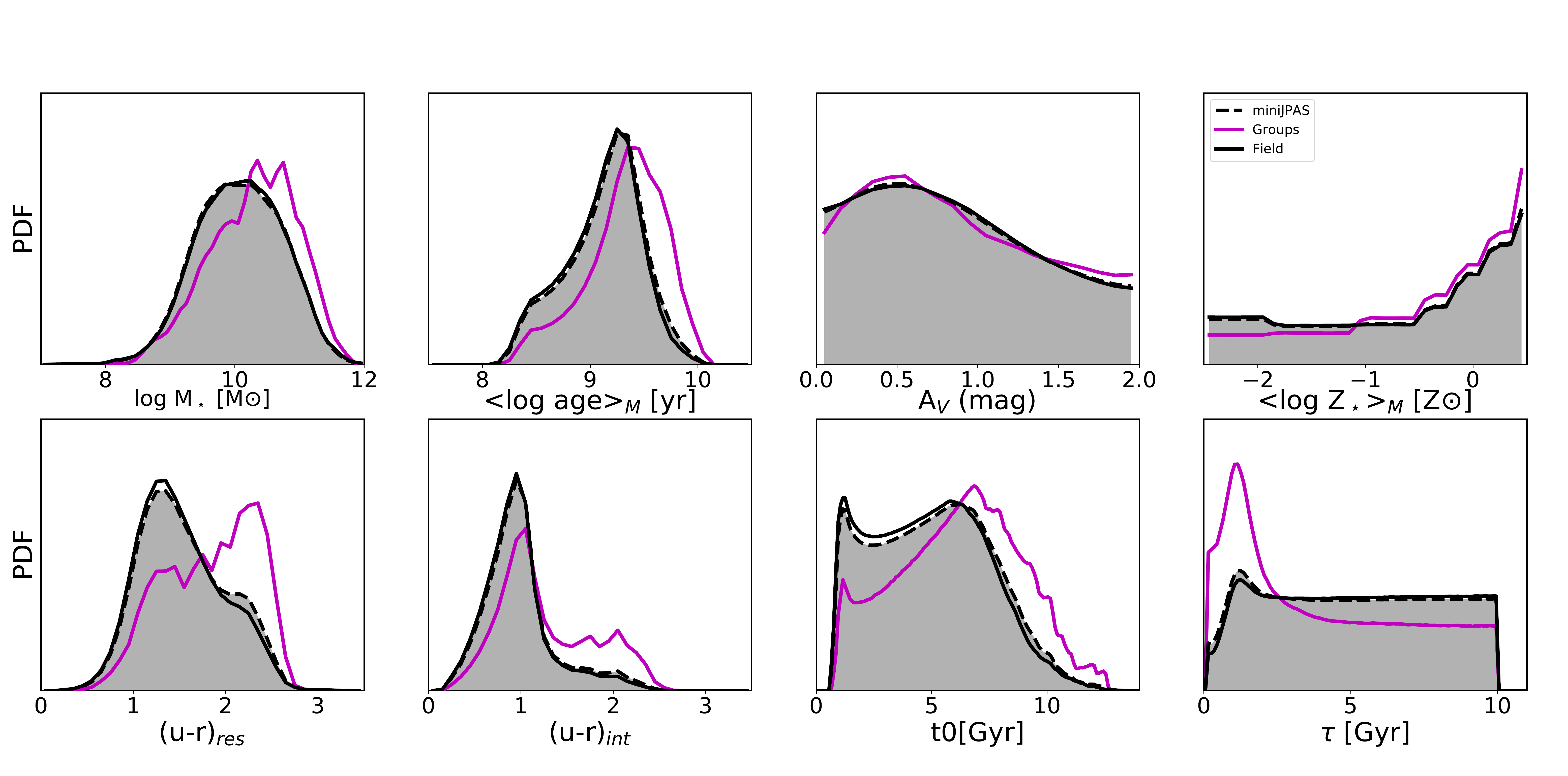}
\caption{Stellar population properties derived from the \js \ (\magauto ) fits with \baysea{} for the whole \mjp \ sample (grey area and dashed black line), for galaxies in the field (black-line), and galaxies in group environment (magenta line). {\it From left to right, and from top to bottom}: PDF of the galaxy stellar mass, mass-weighted ages, stellar extinction, stellar metallicity, rest-frame $(u-r)$ colour, extinction corrected rest-frame $(u-r)$ colour, and the parameters of the SFH: \tc{} and \ta{}.}
\label{fig:hist_SP}
\end{figure*}

 Previous works have pointed out several divergent and/or contradictory results when the properties of galaxies are studied as a function of the environment. For example, SDSS earlier works  found that there is a correlation between galaxy colour, ages, metallicity and SFR with the environment density \citep[e.g.][]{blanton2005}. At a fixed stellar mass, both the star formation and the nuclear activity depend strongly on the local density \citep{kauffmann2004}. On the other hand, \citet{blanton2009} found that the position of the blue cloud and red sequence are independent of the environment. The colour-mass and the colour-concentration indices do not vary strongly with environment \citep{baldry2006}. Further, \citet{bamford2009}, using morphological classification from the Galaxy Zoo, conclude that morphology does not depend on environment once the colour of a galaxy is fixed. However, old galaxies are preferentially located in dense regions, and at a fixed Sersic index, the stellar population ages depend strongly on density \citep{baldry2006}. 
 
 Thanks to the development of codes to find out groups \citep[e.g.][]{yang2007}, galaxy properties have been studied by distinguishing between the satellite and central galaxy populations. For example, \citet{pasquali2010} found that  satellite galaxies are older and more metal-rich than centrals of the same stellar mass. Further,  the slopes of the age-stellar mass  and the metallicity-stellar mass relation become shallower in dense environments \citep{petropoulou2012}.
 In contrast, more recent works using samples at higher redshift ($z <$1) find that the differences between the properties of central and satellites populations are not significant \citep{sobral2021}, although the transformation that drives the evolution of the overall galaxy population must occur at a rate two to four times higher in groups than outside of them \citep{kovac2010}.

The purpose of this section is to compare the stellar population properties of the galaxies in groups with those that are in the field by using \mjp{}. Our SED-fitting analysis allows us to derive in a consistent way the SFH parameters ($\tau$ and $t0$), SFR, sSFR, ages, and metallicity of the stellar populations. 
We also derive the rest-frame galaxy colours corrected for dust extinction to segregate the galaxy populations in blue and intrinsically red galaxies and to study the variation of their properties as a function of the group environment. Moreover, we discuss here the evolution since $z=$1 of the stellar populations properties of galaxy group members in comparison with galaxies in the field.

\subsection{Stellar population properties}

To study the role of group environment in the evolution of galaxies, we first compare the stellar populations properties of galaxies in groups (\Pa{} $\geq$0.7) and in the field (\Pa{} $\leq$0.1). Specifically, we compare the galaxy stellar mass, age, extinction, stellar metallicity, rest-frame colour, extinction corrected rest-frame colour, the time of the onset of the star formation, and the e-folding time (\logM{}, \ageM{}, \Av{}, \logZM{}, \ur{}, \urint{}, \tc{}, and \ta{}, respectively). These properties are calculated using \magauto{} magnitudes.  

The distributions of \logM{}, \ageM{}, \ur{}, \urint{}, \tc{}, and \ta{} for galaxies in groups and in the field are different; in contrast the distributions of \Av{} and \logZM{} are similar (Fig.~\ref{fig:hist_SP}). Mass, ages, and colours of the group population are clearly shifted to higher values with respect to the field population; indicating that on average the galaxy populations in group environment are more massive, redder, and older than in the field. There is a shift of
0.24 dex, 0.21 dex, 0.3, to higher \logM{}, \ageM{}, and to redder colours (Table \ref{tab:SPmedian}). 
The distributions of the SFH parameters \tc{} and \ta{} are significantly different. \tc{} is shifted to earlier epochs, and \ta{} to lower values in groups. These are indicators that group galaxies started to form stars earlier and during a shorter period of time than the galaxies in the field. However, these results do not mean that the group galaxy populations are intrinsically more massive, redder and older than the field population, and the results could be more a consequence of a large fraction of red, older and massive galaxies in dense environments than in the field. We discuss later this point. 

\begin{table*}
    \centering
    \caption{The median and dispersion of the stellar population properties and SFH parameters of the red and blue galaxies in groups and in the field. Last column presents the properties of the BGG}
    \begin{tabular}{l r r r r r r r}
    \hline
    \hline
        SP                             & RG group       &  RG field    &  BG group    & BG field    & Group         & Field  & BGG \\
     \hline
         $\log M_\star$ [$M_\odot$]           &10.78$\pm$0.53& 10.83$\pm$0.58 &  10.16$\pm$0.63  & 10.05$\pm$0.63 & 10.33$\pm$0.66 &  10.09$\pm$0.65 & 11.12$\pm$0.47 \\
         <$\log \mathrm{age}$>$_\mathrm{M}$ [yr]            &9.71$\pm$0.12 & 9.67$\pm$0.13  &  9.26$\pm$0.23   &  9.12$\pm$0.20 &  9.36$\pm$0.28 &   9.15$\pm$0.23 & 9.59$\pm$0.23 \\
         \ur \ [mag]     & 2.36$\pm$0.15  & 2.36$\pm$0.19 &  1.68$\pm$0.43   & 1.48$\pm$0.40  &   1.83$\pm$0.48  &   1.52$\pm$0.43 & 2.34$\pm$0.35 \\
         \urint \ [mag]  & 2.04$\pm$0.19  & 2.01$\pm$0.18&  1.07$\pm$0.26   & 0.93$\pm$0.20  &   1.28$\pm$0.48  &   0.98$\pm$0.30 & 1.82$\pm$0.49\\
         $A_V$ [mag]                   & 0.5 $\pm$0.24  & 0.54$\pm$0.30& 0.96$\pm$0.44  &   0.86$\pm$0.43  &   0.86$\pm$0.45  &   0.85$\pm$0.43 & 0.66$\pm$0.44 \\
         <$\log Z_\star$>$_\mathrm{M}$ [$Z_\odot$] & -0.05$\pm$0.49 & -0.04$\pm$0.51  & -0.37$\pm$0.61 &  -0.48$\pm$0.58  &  -0.30$\pm$0.60  &  -0.46$\pm$0.59 & 0.02$\pm$0.46\\
         $t_0$ [Gyr]                   &7.7$\pm$1.43   & 7.2$\pm$1.54  & 5.89$\pm$1.86 & 4.79$\pm$1.53  &   6.29$\pm$1.93  &   4.90$\pm$1.61 & 6.86$\pm$1.75 \\
         $\tau$ [Gyr]                  &0.95$\pm$0.36  & 0.92$\pm$0.30 & 4.6$\pm$1.76 & 5.3$\pm$1.30   &   3.82$\pm$2.20  &   5.10$\pm$1.58 & 1.10$\pm$1.94\\
         $\tau$/$t_0$                  &0.13$\pm$0.05  & 0.14$\pm$0.04 & 0.92$\pm$0.71& 1.22$\pm$0.75  & 0.68$\pm$0.72  &   1.19$\pm$0.77 & 0.17$\pm$0.34\\
         $\log sSFR$ [Gyr$^{-1}$]         & -1.59$\pm$0.62 & -1.46$\pm$0.75 & -0.31$\pm$0.37 & -0.15$\pm$0.26 &  -0.45$\pm$0.72  &  -0.16$\pm$0.47  & -1.22$\pm$0.90  \\
           \hline
    \end{tabular}
    
    \label{tab:SPmedian}
\end{table*}


\subsection{Identification of red and blue galaxies}
\label{sec:identification}

\begin{figure*}
\centering
\includegraphics[width=\textwidth]{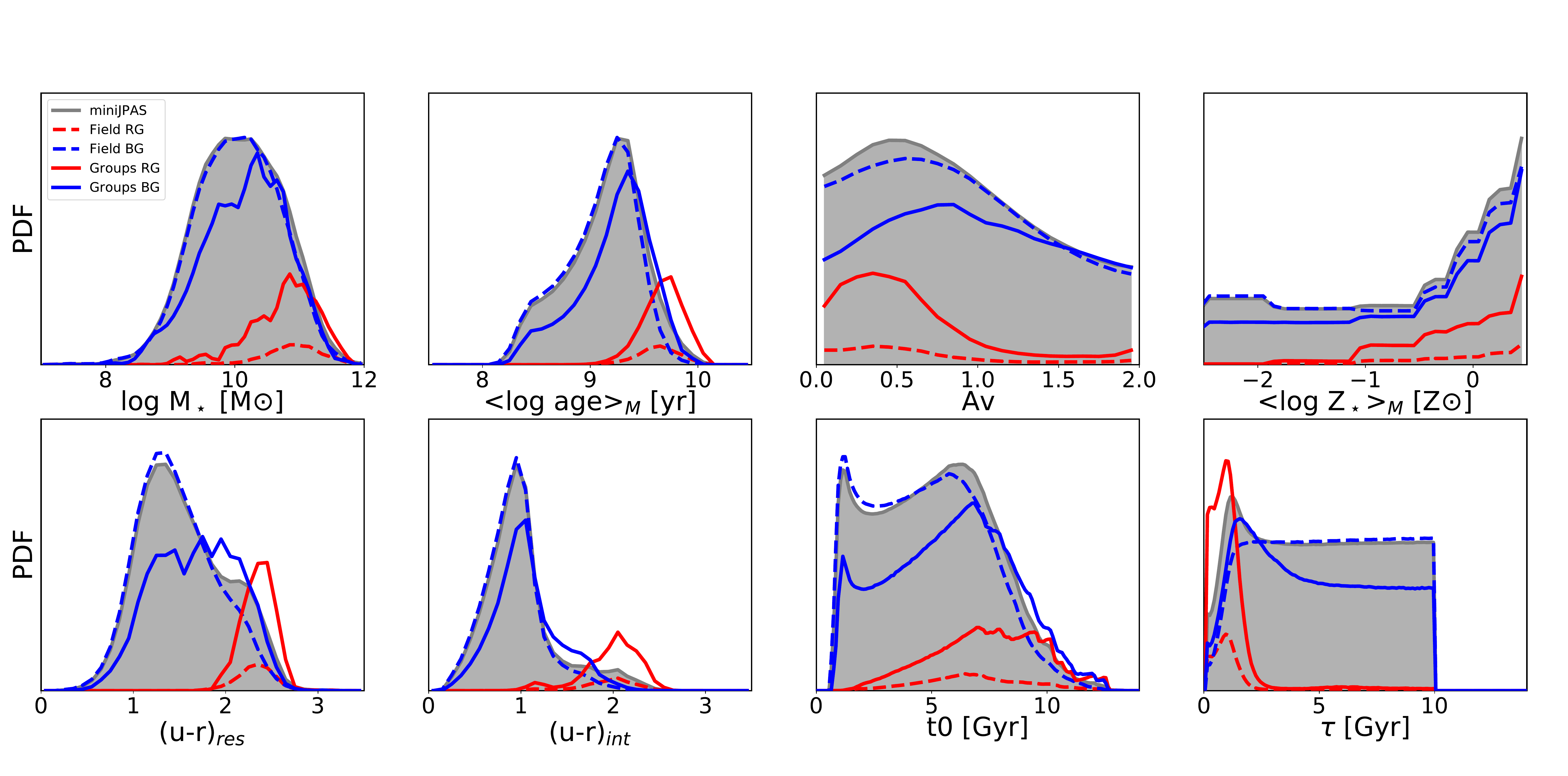}
\caption{PDF distributions for the stellar population properties of the galaxies in \mjp \ (grey area), and red (RG) and blue galaxy (BG) populations in group environment (continuum red and blue lines), and galaxies in the field (dashed red and blue lines).
}
\label{fig:PDFbluered}
\end{figure*}

To identify the red and blue galaxy populations in group and field environments, we use the method developed by \citet{diaz-garcia2019a} using a sample of galaxies at $z < 1$ from the ALHAMBRA survey, that we adapted for the galaxy populations in \mjp{} \citep{gonzalezdelgado21}. We classify galaxies as red or blue according to their extinction corrected rest-frame \urint{}, stellar mass and redshift. We set a colour limit defined by: 

\begin{equation}
\label{eq:redbluegalaxies}
(u - r)_\mathrm{int}^\mathrm{lim} =  0.16 \times (\logMt  \ - 10.) - 0.3 \times (z - 0.1) + 1.7
\end{equation}

\noindent where $z$ is the \photoz\ of the galaxy and \logM{} is its stellar mass. If a galaxy has \urint\ above this ($u - r)_\mathrm{int}^\mathrm{lim}$, it is classified as red; otherwise, the galaxy is labelled as blue. We note that galaxies in the field and in groups are both classified as red or blue using the same criterion detailed in Eq.~(\ref{eq:redbluegalaxies}).

Figure~\ref{fig:PDFbluered} compares the PDF distributions of the stellar population properties of red and blue galaxies in groups environments with the distributions of galaxies in the field. Although the maximum of the PDFs are different, the shape of the PDF of red galaxies in groups is very similar to the PDFs of galaxies in the field. The PDF of blue galaxies in groups is slightly shifted to higher masses, older ages, redder \ur{} and \urint{} colours, and lower \ta\ values with respect to blue galaxies in the field. However, these shifts are very small, with a difference between the median values of $\sim$ 0.1 dex, 0.14 dex, 0.2 mag, 0.14 mag, 0.1 mag, -0.1 dex, for $\logMt$, $\ageMt$, \ur, \urint, \Av, and $\logZMt$\, respectively (see Table \ref{tab:SPmedian}). 

For red galaxies, the shapes of PDFs are very similar, and there is not a significant shift between the median values of the properties with respect to the red galaxy population in the field. The most relevant difference is that  the maximum of the PDF peak is higher for the red galaxies in groups. This is an indication that the fraction of red galaxies is larger in group environment that in the field. This is a well-known result in galaxy clusters/groups studies \citep{dressler1980, balogh2004, balogh2009}.
We will discuss further this point in Sect.~\ref{sec:fraction-quenched}.


\subsection{Specific star formation rate of \mjp{} galaxies}
\label{sec:sSFR}

\begin{figure}
\centering
\includegraphics[width=0.49\textwidth]{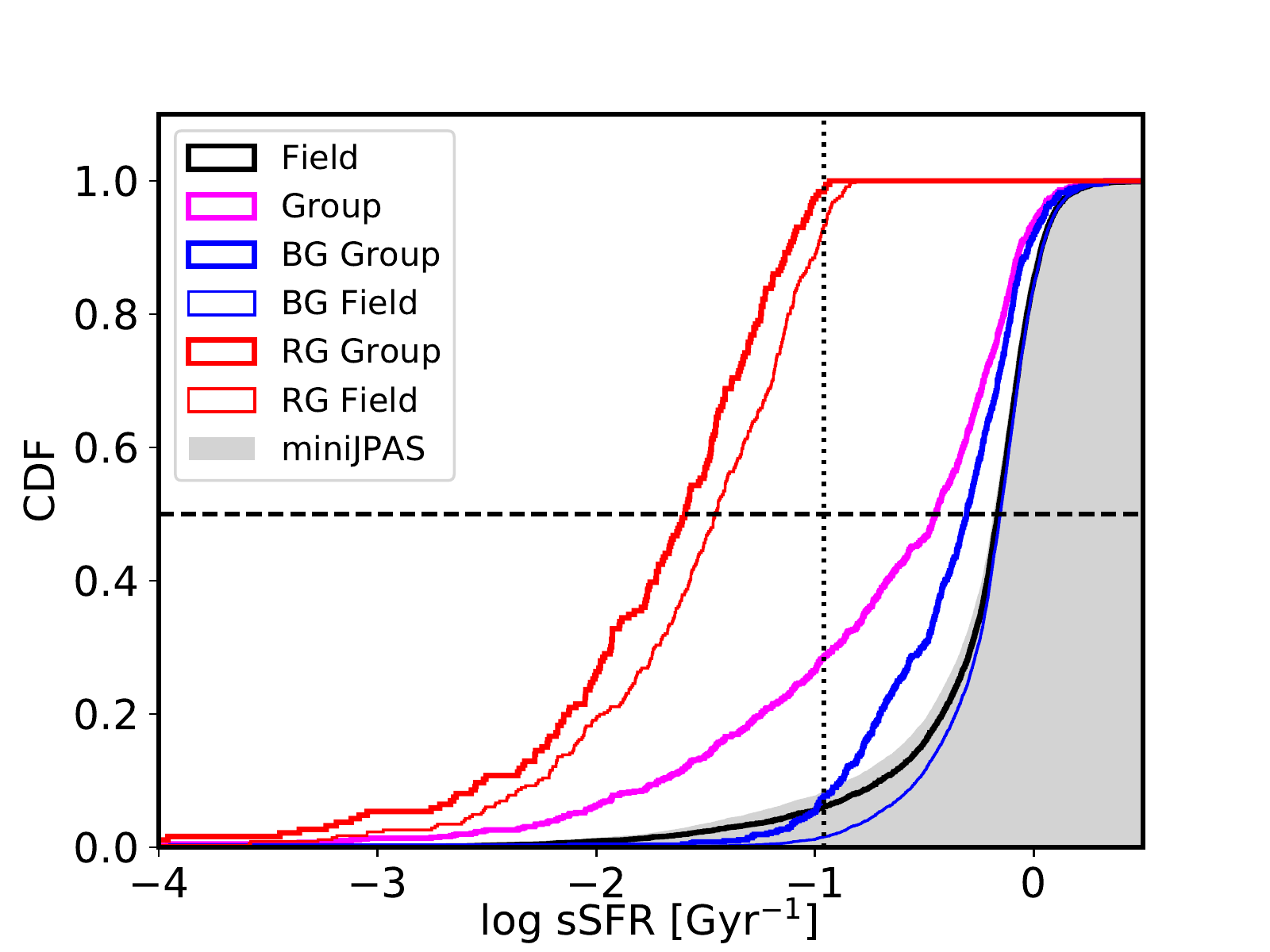}
\caption{Cumulative distribution of the specific SFR of the whole sample (light gray), the field galaxies (black line), the galaxies members of the AMICO groups (magenta), red galaxies in groups (dark red) and in the field (light red), and blue galaxies in groups (dark blue) and in the field (light blue). Vertical line shows the sSFR limit below which the galaxies are considered to be quenched.
}
\label{fig:sSFR}
\end{figure}

The dependence of the SFR and sSFR with environments has been also studied in the past. \citet{kauffmann2004}, for instance, found that sSFR is the most sensitive property to the local galaxy density. In contrast, other works suggest that the sSFR and its relation with the galaxy stellar mass of star forming galaxies is independent of the environment up to $z\sim$1 \citep{peng2010, darvish2016, sobral2021}. However,  in very dense environments,  it was found a reduction in the SFR \citep{haines2013}; although it could be produced by the presence of a large fraction of red-disk galaxies with respect to less dense  environment \citep{erfanianfar2016}.

We calculate the SFR of each galaxy using the SFH derived from the fits by adding the mass gained during the last 100 Myr, and dividing this quantity by this period of time. This number is representative of the current SFR in the galaxy, and it is nearly equal to the SFR calculated using a period of time $\sim 30$~Myr,  which is the epoch in which the galaxy optical luminosity is dominated by O and B0 stars, and the H$\alpha$ line is detected in emission \citep{asari2007}. 

Figure~\ref{fig:sSFR} shows the cumulative distribution of sSFR values for galaxies in \mjp{}, and in the subsamples of galaxies in groups and in the field. Clearly, there is a difference of $\sim -0.29$~dex (median value) between the galaxies in groups with respect to those in the field. Thus, as it was pointed out by \citet{kauffmann2004}, sSFR is very sensitive to the environment, being lower in environments of high local density. However, this shift to lower sSFR is mainly due to the existence of a larger fraction of red galaxies in groups than in the field.
The difference is smaller when the comparison is done considering separately the red and blue galaxy populations. Thus, the difference in the sSFR between galaxies in groups and in field is $-0.16$ and $-0.13$~dex for the blue and red populations, respectively. Therefore, we conclude that in the \mjp{} sample, the dependence of the sSFR with the group environment is small, $\sim 0.15$~dex, when the red and blue galaxy populations are considered separately.


\subsection{$M_\star$--sSFR relation in groups}
\label{sec:Mainsequence}

\begin{figure}
\centering
\includegraphics[width=0.49\textwidth]{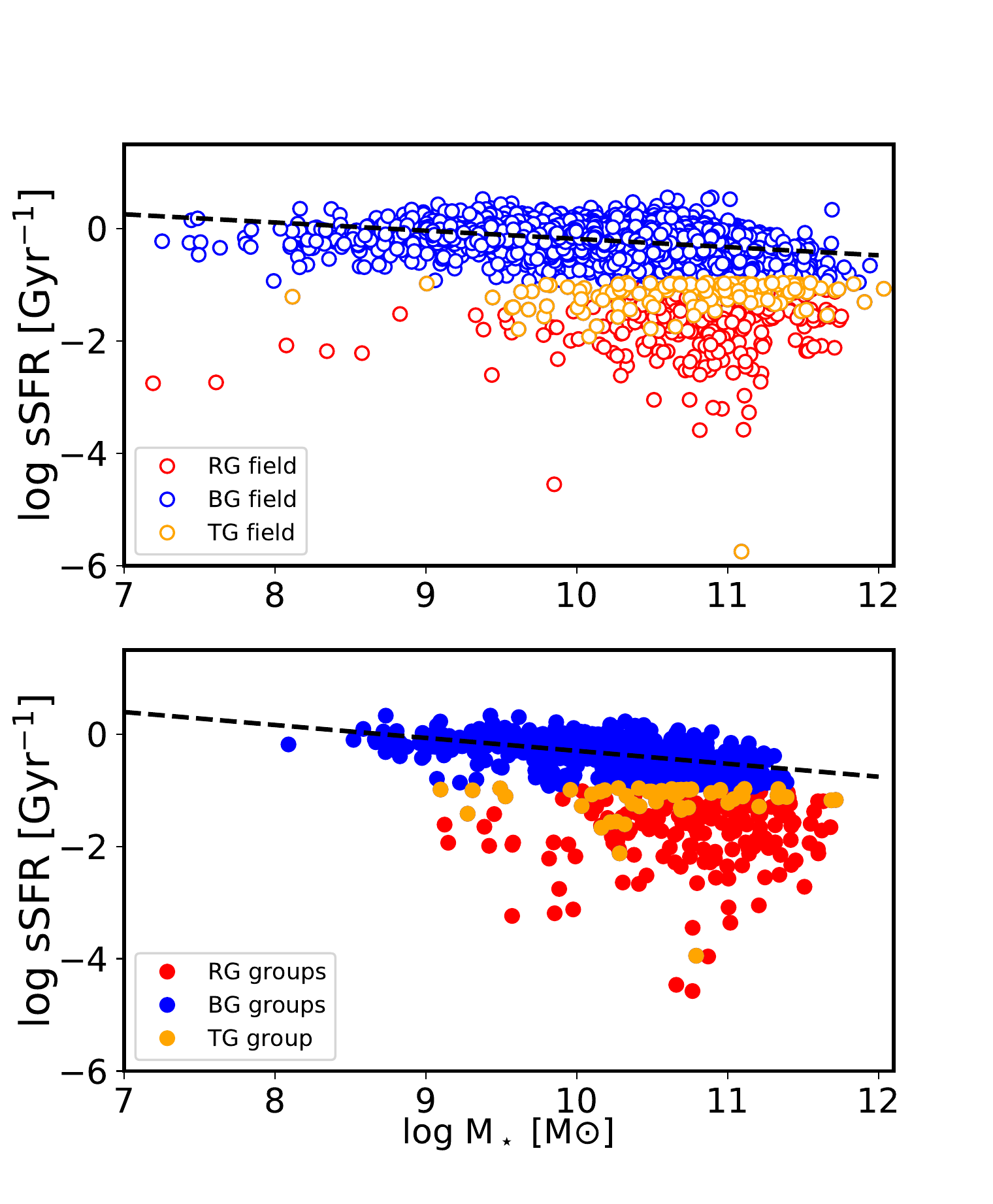}
\caption{$\log sSFR-$\logM{} relation for galaxies in the field (upper panel) and group environment (bottom panel). Red and blue galaxies are represented by blue and red circles, respectively. Blue galaxies that are in a transition phase (see Sec.\ref{sec:transition}) are represented by orange circles. Black dashed lines represent the fit obtained for the SFMS in each set of galaxies.}
\label{fig:SFMS}
\end{figure}

The star forming main sequence (SFMS) is the correlation between the SFR of a galaxy and its galaxy stellar mass. 
This relation is tight with a dispersion of only 0.2-0.3 dex at a fixed stellar mass and with a slope that is close to, but below 1 \citep{brinchmann2004, renzini2015, peng2010}. 
There is also a tight relation between the intensity of the star formation and the stellar mass density of each galaxy region \citep{gonzalezdelgado2016, cano2016}, that defines a local SFMS, with a slope similar to the global galaxy SFMS. This suggests that local processes are relevant to determine the SFR in the disk of galaxies, probably through a density dependence of the SFR law \citep{gonzalezdelgado2016}. Recent results from the MaNGA survey confirm that the star formation in galaxies is governed by local processes within each spaxel \citep{bluck2020}.

Nowadays, it is well-known that the star formation happening in the Universe since $z\sim 1$ is mostly produced within blue galaxies, which in turn result in the SFMS \citep{brinchmann2004, madau14}, and the SFMS exists since high-redshift \citep{noeske2007, elbaz2007, whitaker2012, tasca2015}. A similar correlation exists between the sSFR and the galaxy stellar mass. Because the SFMS shows a slope $<1$, the sSFR declines weakly with increasing mass \citep{salim2007, schiminovich2007}. 

Many previous works have found that the SFMS is independent of the environment \citep{peng2010, darvish2015, darvish2016}, and have shown that the dependence of the SFR and the sSFR in the environment may be due to a larger fraction of quiescent galaxies in high density environments. They suggest that environment does not regulate the build-up of mass in star forming galaxies. 

Figure~\ref{fig:SFMS} shows the relation between log sSFR and \logM{} for galaxies in low density environments and in groups. Blue (BG) and red galaxies (RG) are properly separated in the log sSFR and \logM{} relation. As expected, BGs define the SFMS at a fixed stellar mass, while RGs are below this observational relation.
To explore the dependence of the SFMS with the environment, we fit a linear relation (log sSFR = b + a $\logMt{}$) only to the star forming galaxies. \citet{peng2010} proposed that only blue galaxies with sSFR >  0.1 Gyr$^{-1}$ are actually star forming galaxies. Here, we exclude all the RGs and BGs that have sSFR below this threshold for fitting the SFMS. The results of the fit are $(a, b) = (-0.23\pm0.02, 2.0\pm0.2)$ for galaxies in groups, and $(a, b) = (-0.15\pm0.01, 1.3\pm0.1)$ for galaxies in the field. The differences between SFMS of star forming galaxies in groups and in the low density environment are small, being negligible in the low mass bins and $\sim -0.18$~dex in log sSFR for $\logMt{} >$ 11. Thus, there is a reduction of star formation only in massive galaxies that are in the group environments with respect to the galaxies in less dense environments.


\subsection{Mass--colour diagram vs. environment}
\label{sec:mass-colour}

\begin{figure*}
\centering
\includegraphics[width=\textwidth]{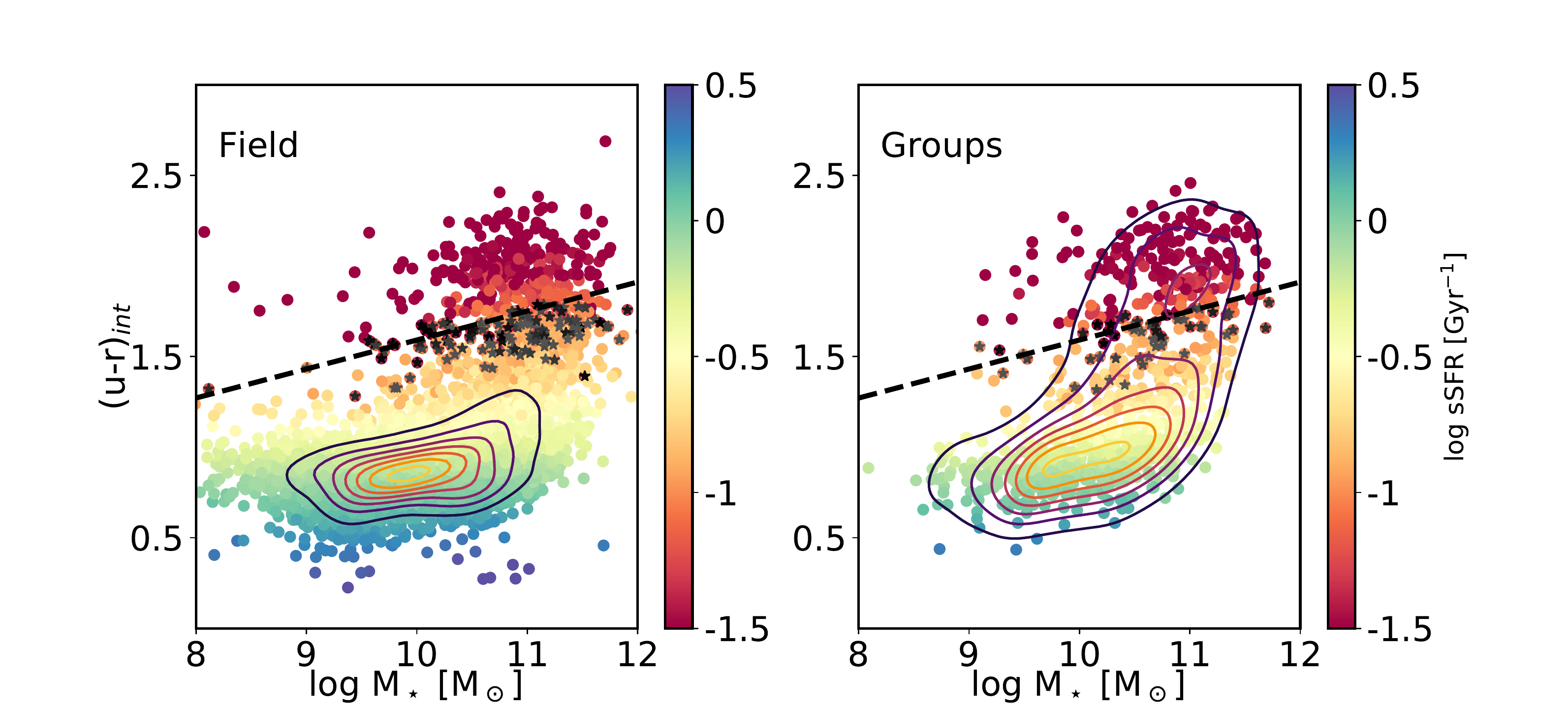}
\caption{Mass-colour (rest-frame $(u-r)$,  corrected for extinction) for the field galaxy population (left panel) and galaxies in group environment (right panel). Dashed lines show the $(u-r)^{lim}_{int}$ for the mean redshift of the galaxy population, to separate blue and red galaxies. Grey stars represent the blue galaxies that are considered to be in a transition phase (see Sec.\ref{sec:transition}).The contours represent the density distribution of points in the colour-Mass plane.
}
\label{fig:Masscolour}
\end{figure*}

The bimodal colour distribution of  galaxy populations is clearly seen in the \urint{}-\logM{} diagram (Fig.~\ref{fig:Masscolour}).The rest-frame colour corrected for extinction (e.g.~\urint{}) is more useful than \ur{} to segregate the red and blue populations and to account for the fraction of red and star forming galaxies in the sample. The \mjp{} galaxies in the \logM--\urint{} diagram are clearly distributed in the red sequence and the blue cloud, with the galaxies in the red sequence being typically old and metal-rich \citep{gonzalezdelgado21}. This  mass-colour bimodal distribution is in place for the group and the field galaxy populations, although the fraction of galaxies that populate the red sequence and the blue cloud are different (Fig.~\ref{fig:Masscolour}). 

It is well-known that the bimodal colour distribution of galaxies is connected to the SFR and sSFR of galaxies, in the sense that blue galaxies have higher SFR and sSFR than red galaxies \citep{brinchmann2004, salim2007, renzini2015, gonzalezdelgado2016, gonzalezdelgado2017, lopezfernandez2018}.  
This result is clearly confirmed by our analysis in the AEGIS field
in the colour \urint{} - \logM{} diagram (Fig.~\ref{fig:Masscolour}) for galaxies in the field and in groups. We stress that in both diagrams, red galaxies (located above the dashed line in  Fig.~\ref{fig:Masscolour}) are redder than their $(u-r)^{lim}$, calculated with Eq.~(\ref{eq:redbluegalaxies}) for each galaxy,  and are  characterised by a sSFR (SFR/$M_\star$) $<$ 0.1 Gyr$^{-1}$, whereas those in the blue cloud usually have sSFR $>$ 0.1 Gyr$^{-1}$.

Red colours and a sSFR below a given threshold are used as proxy for identifying quenched galaxies. Here, we find that the two proxies provide a similar fraction of quenched galaxies. Using as threshold $sSFR<0.1$~Gyr$^{-1}$, we find that $28$~\% and $6$~\% of the galaxies in groups and in the field are quenched, respectively. Using the \urint{} colour--mass relation, the fraction of quenched galaxies are $23$~\% and $5$~\% in groups and in the field, respectively. Thus, the two proxies yield similar results, and indicate that there is in average a $20$~\% excess of quenched galaxies in dense environments, in agreement with \citet{balogh2009} that studied the fraction of red galaxies in groups at $z\sim 0.4$.
Furthermore, we find that in average the fraction of quenched galaxy population is significantly higher in groups than in less dense environments, and it is also significantly higher than in the whole AEGIS galaxy population (lower than $8$~\%).
Note, however, that this excess is a function of \logM{} and  redshift. We discuss later this point (see Sec. \ref{sec:FractionRedBlue}).
\subsection{Properties of the group central galaxies}
\label{sec:BGG_properties}

\begin{figure*}
\centering
\includegraphics[width=\textwidth]{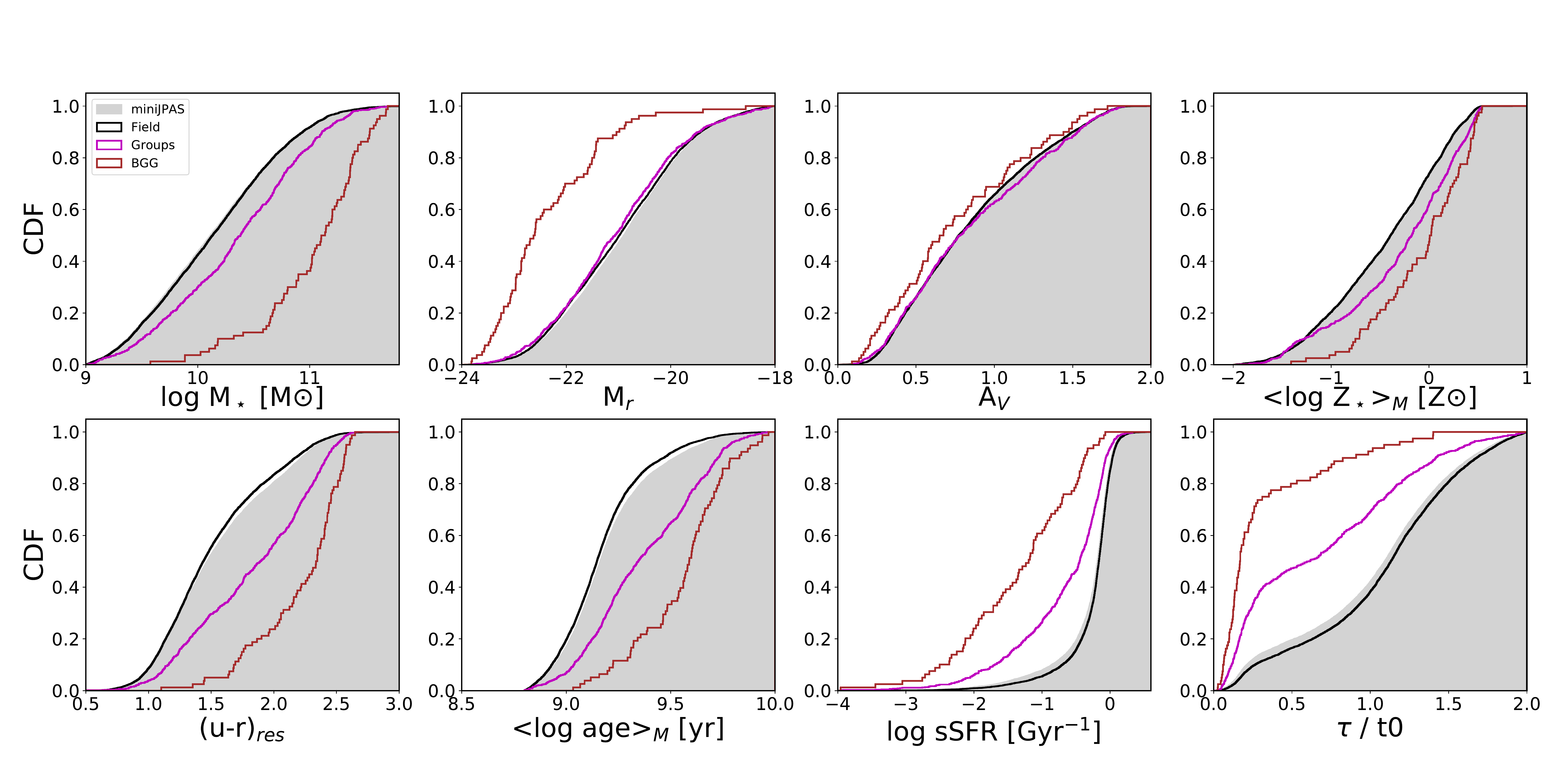}
\caption{Distribution of the properties of the central galaxy (brown line), compared with the properties of galaxies in groups (magenta line) and galaxies in the field (black line). {\it From left to right, and top to bottom:} stellar mass; absolute magnitude in the $r$ band; intrinsic stellar extinction; stellar metallicity; rest-frame $(u-r)$ colour; stellar age (mass-weighted); sSFR; and ratio of the SFH papameters t0 and $\tau$. 
}
\label{fig:histBCG}
\end{figure*}
Here, we refer as central galaxies to the most massive and brightest galaxies in each of the AMICO groups detected. As we have explained in Sect.~\ref{sec:analysis}, they are the most massive galaxies close to the group center. 
We now study the stellar population properties of the BGGs (see Fig.~\ref{fig:histBCG}), and we compare them  with the properties of the other  members of the groups, and with the sample of galaxies in the field environment. 
BGGs are significantly more massive ($\sim$ 1 dex), brighter ($\sim$ 1.6 mag), redder (\ur{} $\sim$0.9 higher), older (\ageM{} $\sim$0.9 dex higher), and more metal rich (\logZM{} $\sim 0.4$~dex higher) than the rest of the galaxy population in \mjp{}. BGGs are slightly less affected by extinction (\Av{} $\sim$0.14 mag lower) that the rest of the galaxy sample. In terms of their star formation activity, BGGs are the galaxies with the lowest sSFR,  $\sim 1$~dex below the rest of the galaxy population in \mjp, suggesting that the star formation has been shut down significantly in these galaxies. Additional evidence of this shut down of the star formation taking place a long time ago and/or happening in a short period of time comes from \tato{}, which is very small ($\sim$0.17) in comparison with the general \mjp{} galaxy population (\tato{} $\sim 1.3$).
Differences between BGGs and the other  group members are also significant, being more massive, more luminous, more metal rich, older and with lower sSFR than the rest of the galaxies in groups (see Table \ref{tab:SPmedian}). The median values of sSFR, \ur{} and \tato{} suggest that BGGs are red quiescent galaxies; however, there are $\sim 38$~\% of the BGGs that are still forming stars with $sSFR > 0.1$~Gyr$^{-1}$.
This fraction, though, decreases with redshift. Only $\sim$20$\%$ of the BGGs at $z <$0.3 have $sSFR > 0.1$~Gyr$^{-1}$; thus $\sim$80$\%$ are quiescent galaxies. This is in agreement with previous results from SDSS that show that 80$\%$ of the central galaxies in clusters at z$<$0.1  have ceased their star formation, independently of their stellar mass \citep{vonderlinden2010}.

\subsection{The evolution of the stellar population properties}
\label{sec:characterization}

\begin{figure}
\centering
\includegraphics[width=0.49\textwidth]{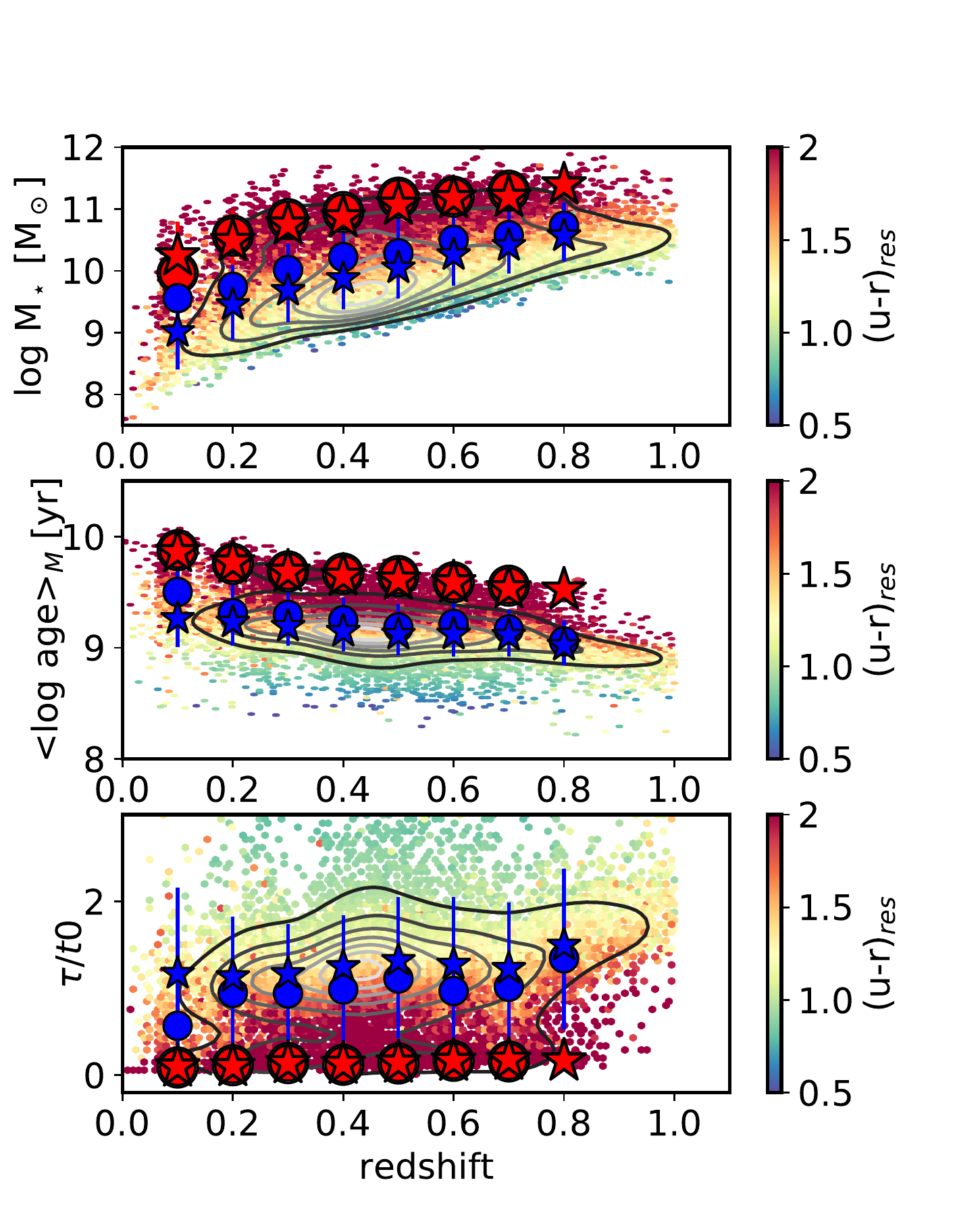}
\caption{Evolution of the stellar mass, age and the ratio of the SFH parameter $\tau$/$t0$. The contours represent the density distribution of points. The dots  represent the average values of each property in each redshift bin. Blue and red circles (stars) are the values for blue and red galaxies in group (field) environments. The dispersion with respect to the average are shown as error bars.  
}
\label{fig:redshiftSP}
\end{figure}

The environment can play a different role at different epochs. Here, we explore the properties of the red and the blue populations in groups and in low density environment as a function of redshift. In particular, we explore the evolution of \logM{}, \ageM{}, and  $\tau$/$t0$. 

Previously, we have discussed the evolution of the \mjp{} red and blue galaxy populations with redshift \citep{gonzalezdelgado21}. We have found that red and blue galaxies are properly distinguished by their stellar content and properties. At any redshift bins below $z=1$, the red galaxies are older and redder than the blue galaxies; and both galaxy populations are ageing since $z=1$. The red galaxies are also more massive than the blue population. The median of the stellar mass values in our sample are higher at $z=1$ than at $z=0$. However, this is a consequence of the incompleteness of the sample, because galaxies less massive than $10^{10}$~$M_\odot$ are not detected at $z> 0.8$, and galaxies with $2\times 10^{8}$~$M_\odot$ are detected only up to $z=0.15$  \citep[see Fig.~19 in][]{gonzalezdelgado21}.This  also applies to the actual sample.

Figure~\ref{fig:redshiftSP} shows the average value of the different galaxy properties in each redshift bin. Blue and red galaxies are properly distinguished by their stellar content at any redshift bin. The local density of galaxies is not playing a relevant role to set the average properties of the red galaxies, because galaxies in groups and in the field have in average equal \logM{}, \ageM{}, and $\tau/t_0$ at any epoch; and similarly with blue galaxies, although at any redshift, blue galaxies in groups are slightly more massive, and the star formation 
extends during shorter period of time. It is worth mentioning that the small differences between blue galaxies in groups and in the field are almost constant, and independent of the redshift, although they tend to increase at lower redshifts.
This is an indication that there is a larger fraction of blue galaxies in the transition phase to be transformed in red galaxies in groups. We discuss this point in Sect.~\ref{sec:transition}.



\section{Fraction of red and blue galaxies vs. environment}
\label{sec:fraction-quenched}

\subsection{The colour-density relation in \mjp{}}
\label{sec:density}


\begin{figure}
\centering
\includegraphics[width=0.49\textwidth]{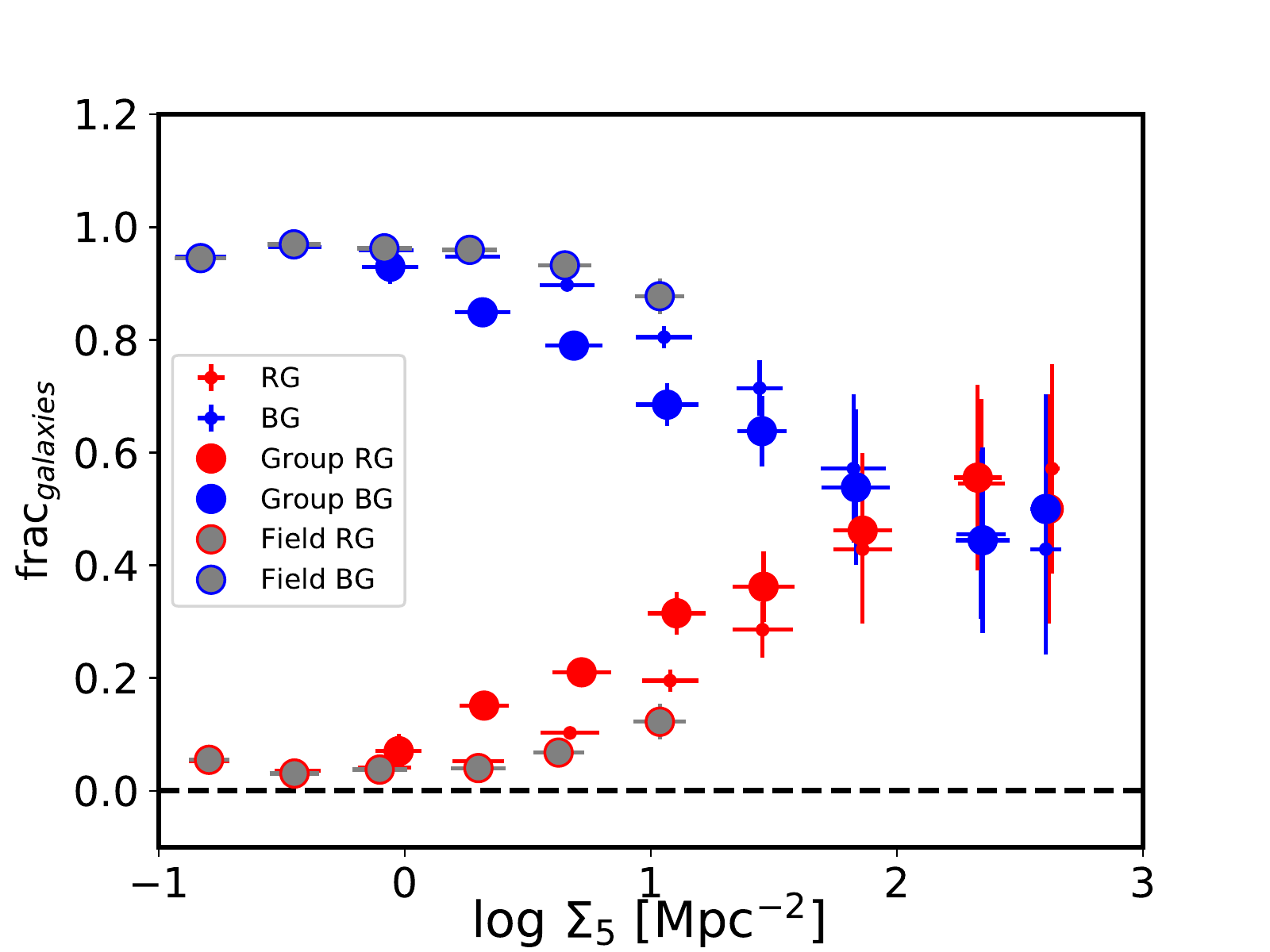}
\caption{ Fraction of red and blue \mjp{} galaxies in different bins of $\log \ \Sigma_5$. The results for the whole galaxy sample (dots), galaxies in groups (red and blue circles), and in the field (grey-red and grey-blue circles) are shown.
}
\label{fig:colour-density}
\end{figure}

Large structures, such as galaxy clusters, have been extensively used to study the role that environment plays in galaxy evolution. In particular, regarding the transformation of late type to early type galaxies, and how this depends on the local density number of galaxies. The pioneering work of \citet{dressler1980} obtained a clear morphology-density ($T-\Sigma$) relation, showing an increase of the fraction of early type galaxies as a function of the local density number of galaxies. 

The \mjp{} has proven to be very successful survey for detecting clusters and groups with masses down to $10^{13}$~$M_\odot$ (Maturi  et al. in prep.). 
Here, we show that \mjp{}  and our approach based on the bimodal colour distribution of galaxies is valid to study the role that group environment plays in quenching the star formation in galaxies. Firstly, we show that our analysis can reproduce a relation similar to the morphology-density, $T-\Sigma$, relation by \citet{dressler1980}, but using the colour blue-red classification of the sample instead of the morphology of the galaxy. This is justified because it is well-known that the separation of the galaxy populations in the red and blue galaxies in the colour-mass diagram is well correlated with  the stellar population properties of the galaxies \citep{kauffmann2003a, kauffmann2003b}, and also with their morphology. In the local Universe, red galaxies are mainly elliptical or spheroidal dominated systems with little star formation, while blue galaxies are disk dominated systems with still ongoing star formation mostly concentrated in their spiral arms \citep{blanton2009}. On the other hand, previous works have  confirmed a colour-density relation 
\citep[e.g.][]{lewis2002, kauffmann2004, rojas2005, weinmann2006, liu2015, moorman2016}.

Figure~\ref{fig:colour-density} shows the fraction of red ($f_\mathrm{R}$) and blue ($f_\mathrm{B}$) galaxies in \mjp{} as a function of the local number density of galaxies measured by \s{}. 
The error bars associated to $f_\mathrm{R}$ and $f_\mathrm{B}$ in each bin of \s{} are estimated as the confidence intervals of a binomial distribution. We use the normal approximation and a 68$\%$ confidence level. This distribution is characterized for giving equal confidence intervals, thus errors, for $f_\mathrm{R}$ and $f_\mathrm{B}$. 
These assumptions are also applied to calculate the error in Fig. \ref{fig:fracredblue}, and Fig. \ref{fig:butcher}. For figures in Sec. \ref{sec:Discussion},  the error bars are derived after propagation of the confidence intervals associated to the different fraction of galaxies involved in the calculation of the property shown in the figure.

$f_\mathrm{R}$ increases with \s{}, while $f_\mathrm{B}$ decreases. 
For example, $f_\mathrm{R}$ increases from 0.04 at a value of \su{} = 0.02, which is more representative of a field environment, up to 0.2 when \su{} = 0.9, a value representative of groups. This increase is more significant, up to $\sim 65$~\%, when sampling the big-structure of the AEGIS field, the cluster mJPC2470-1771 (Rodr\'iguez-Mart\'in et al., 2022, submitted). It is worth to notice that $f_\mathrm{R}$ in groups is significantly higher than in the field for \su{} $>$0; for example,  $f_\mathrm{R}$ is 0.36 for \su{} = 1.3, while it decreases down to zero for galaxies in the field. We can conclude that the color-density relation in AEGIS is driven by the galaxy group population.

\subsection{Fraction of red and blue galaxies in groups}
\label{sec:FractionRedBlue}

\begin{figure}
\centering
\includegraphics[width=0.49\textwidth]{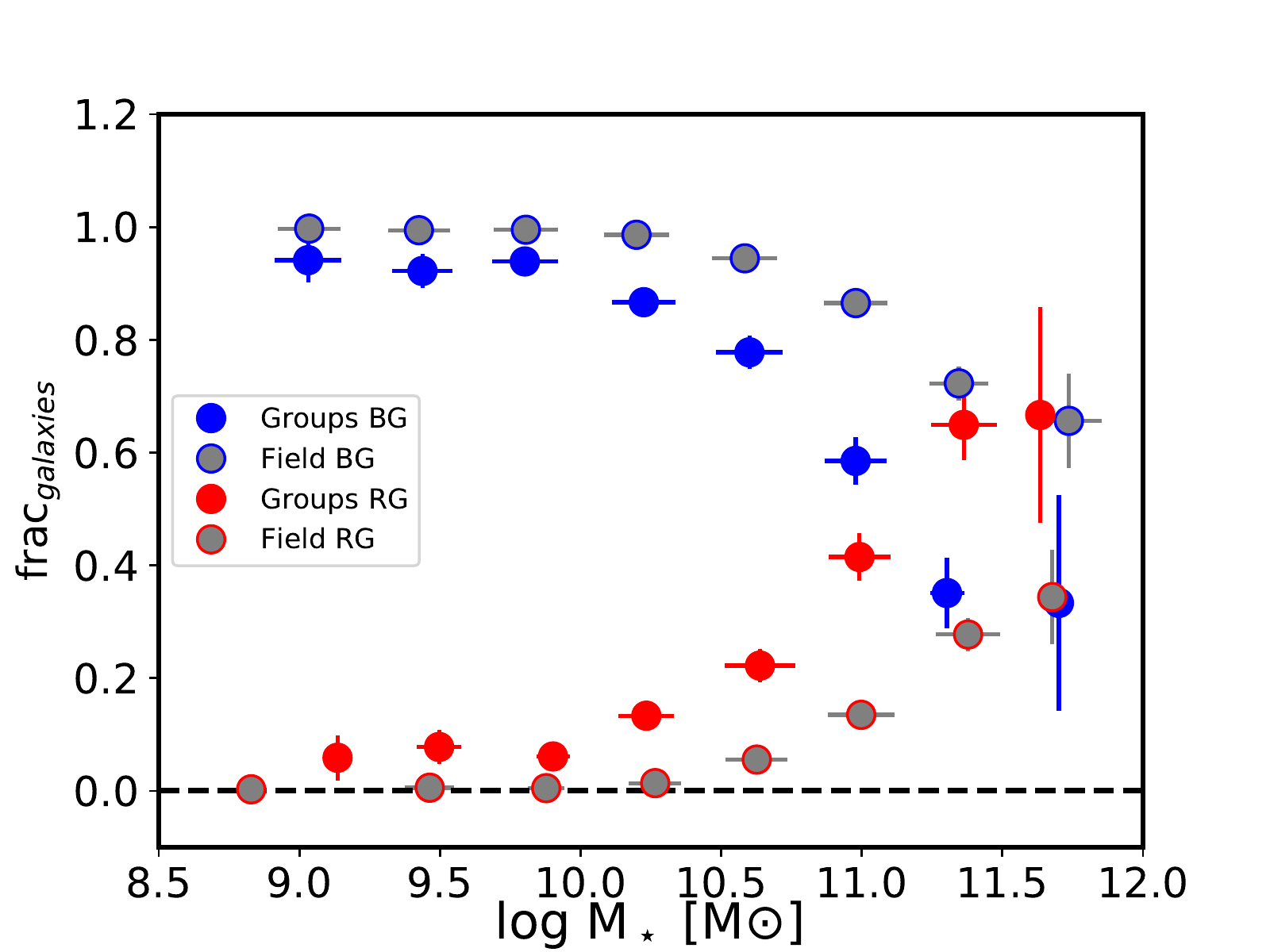}
\caption{Fraction of red and blue galaxies in different stellar mass bins for galaxies in groups (red and blue circles) and in the field (grey-red and grey-blue circles).The bins have a width of 0.4 dex in stellar mass and the points are plotted at the mean mass of the galaxies that belong to each bin.
}
\label{fig:fracredblue}
\end{figure}

This section discusses the impact of stellar mass and group environment on the quenching process. Broadly, two distinct scenarios have been proposed for quenching: mass quenching and environmental quenching. \citet{peng2010} found that in the zCOSMOS sample ($z\leq 1$) the effects of stellar mass and environment on the fraction of star forming and passive galaxies are separable. 
In contrast, other studies based on the CANDELS survey \citep{liu2021} have found that the quiescent fraction is significantly large at the high-mass end and at local environmental overdensities, which suggests a dependence of quenching on both mass and local environment.

Figure~\ref{fig:fracredblue} shows $f_\mathrm{R}$ and $f_\mathrm{B}$ as a function of the galaxy stellar mass for the group and field subsamples.
Notice, that these fractions, $f_\mathrm{R}$ and $f_\mathrm{B}$, are not corrected by volume incompleteness of the sample. In \citet{gonzalezdelgado21} we have shown that galaxies with $\logMt{}$ $\sim$ 10 can be detected up to z$\sim$0.8 in \mjp{}.  However, at the highest redshift bin, we can calculate $f_\mathrm{R}$ only for the bins of mass above 10$^{11}$ M$\odot$. 

Galaxies with $\logMt \leq 9$ were not detected in groups, although they are in the field. For $\logMt > 10$, the fraction of red (blue) galaxies is significantly higher (lower) in groups than in the field. However, the differential effect is significantly higher for $\logMt \geq 11$ than for lower masses. This suggests a dependence of quenching on both mass and group environment. 



\begin{figure}
\centering
\includegraphics[width=0.5\textwidth]{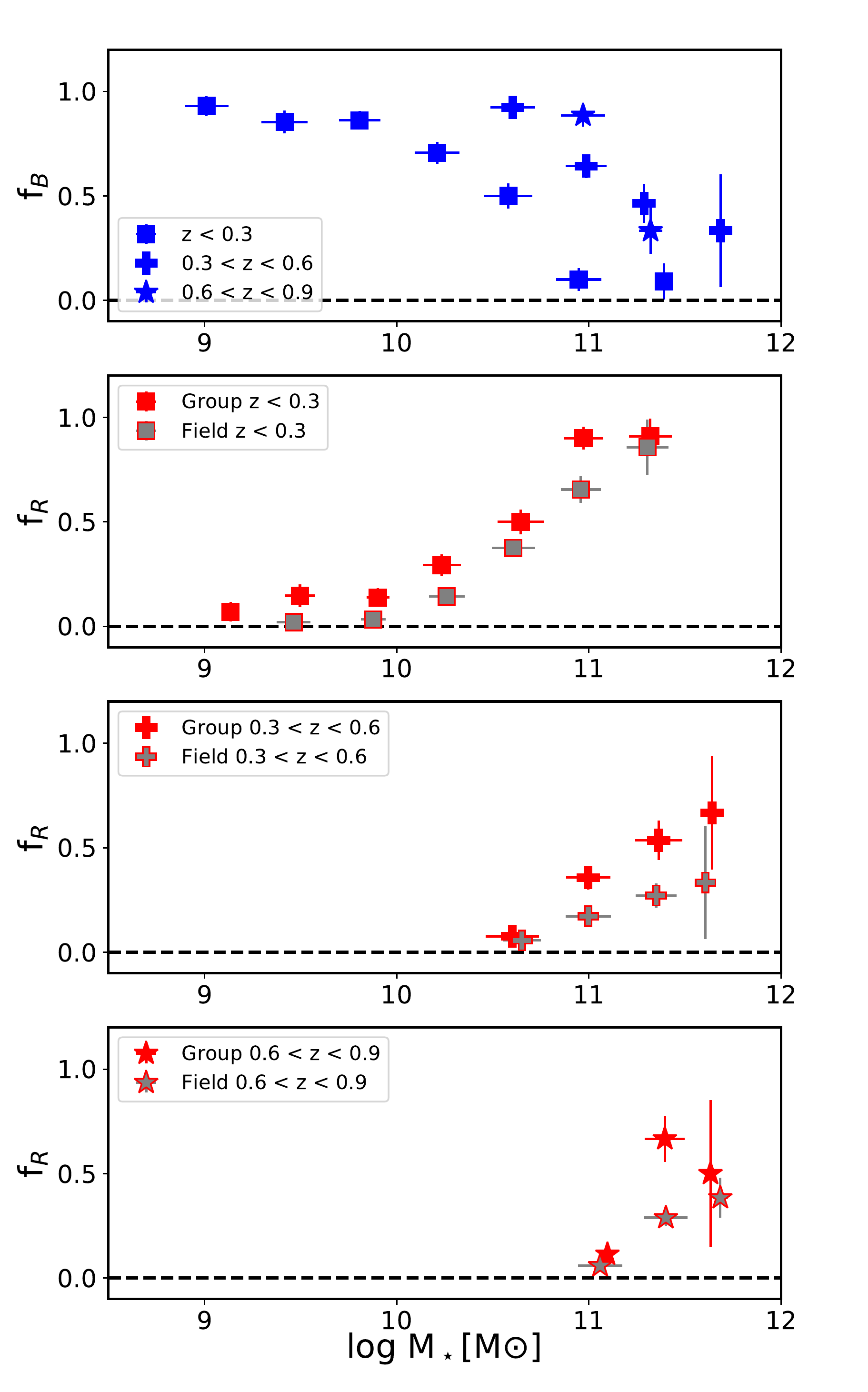}
\caption{Evolution of the fraction of blue and red galaxies as a function of the galaxy stellar mass. Upper panel: Fraction of blue galaxies in groups in three redshift bins, $z < 0.3$ (squares), $0.3 < z < 0.6$ (cross) and $0.6 < z < 0.9$ (stars). Middle upper and bottom panels: Fraction of red galaxies in groups (red symbols) and in the field (grey-red symbols) for three redshift bins.
}
\label{fig:butcher}
\end{figure}

It is well-known that the fraction of blue galaxies in the core of galaxy clusters at intermediate redshift ($z\sim 1$) is higher than in clusters at lower redshift \citep{butcher1984}. To test this result in a less dense environments, we study the evolution of galaxy populations in \mjp{} groups.  We split the sample in three redshift bins ($z\leq 0.3$, $0.3 < z \leq 0.6$, and $0.6 < z \leq 0.9$), and we compare the fraction of red and blue galaxies in each of them. Figure~\ref{fig:butcher} clearly shows that the fraction of blue galaxies in \mjp{} at a fixed \logM{} is higher at the intermediate redshift bins than at $z < 0.3$.  For example, for galaxies with ($\logMt \sim 10.6$), $f_\mathrm{B}$ is $50$~\%, $92$~\%, and $100$~\% for $z\leq 0.3$, $0.3 < z \leq 0.6$, and $0.6 < z \leq 0.9$, respectively. This result confirms the Butcher-Oemler effect in \mjp{}. 

Complementary, the fraction of red galaxies ($f_\mathrm{R}$) in groups evolves with redshift, being higher at lower redshifts. To differentiate the effect of redshift  and stellar mass, we compare $f_\mathrm{R}$ in groups and in the field as a function of \logM{}. It is worth mentioning that the fraction of red galaxies detected in \mjp{} at $0.3 < z < 0.9$ is small, $f_\mathrm{R} \sim 10$~\%, and due to volume incompleteness of the sample, only galaxies with stellar mass above \logM{} $\sim 8.8$ at $z = 0.3$, and $9.9$ at $z= 0.7$ are detected in \mjp{} \citep[see Fig.~19 in][]{gonzalezdelgado21}. We find that the evolution with redshift is significant. For instance, $f_\mathrm{R}$ in groups and for $\logMt \sim 11$ ranges from $0.9$ at $z < 0.3$, to $0.36$ at $0.3 < z \leq 0.6$, and $0.11$ at $0.6 < z < 0.9$. However, the increase in $f_\mathrm{R}$ in groups with respect to the field ($\Delta f_\mathrm{R} = f_\mathrm{R}^\mathrm{G} - f_\mathrm{R}^\mathrm{F}$) does not vary significantly with redshift, where the mean and the standard deviation are $\Delta f_\mathrm{R} = 0.13 \pm 0.06$, $0.12 \pm 0.11$, and $0.14 \pm 0.16$ for the low, intermediate, and higher redshift bins, respectively.
Further, the incremental effect of $\Delta f_\mathrm{R}$ is less dependent on the galaxy mass at $z\leq 0.3$, than at $z > 0.6$. This is also in agreement with \citet{liu2021} that find that the process of star formation quenching exhibits a strong dependence on stellar mass at early epochs, and the mass dependence of quenching tends to decrease with cosmic time.


%
\section{Discussion}
\label{sec:Discussion}

\subsection{Fraction of quenched galaxies in groups}

\label{sec:FractionQuenched}

\begin{figure}
\centering
\includegraphics[width=0.49\textwidth]{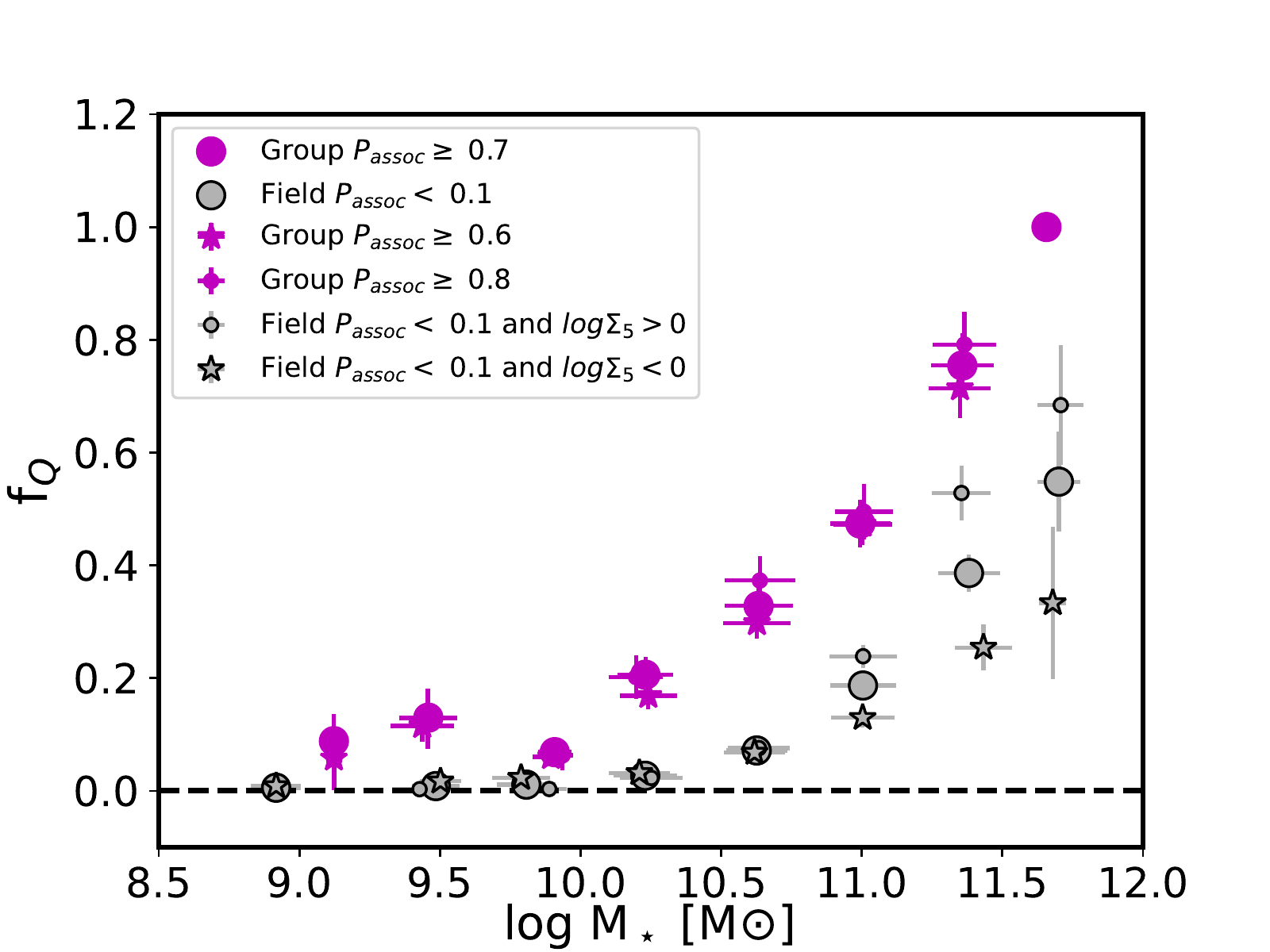}
\caption{Fraction of quenched galaxies in different stellar mass bins for galaxies in group environments and in the field. Different assumptions in $P_\mathrm{assoc}$ are taken for selecting galaxies in groups: $P_\mathrm{assoc} \geq 0.7$, $P_\mathrm{assoc} \geq 0.8$, and $P_\mathrm{assoc} \geq 0.6$ (magenta circles, dots and stars, respectively); and galaxies in the field: $P_\mathrm{assoc} < 0.1$, $P_\mathrm{assoc} < 0.1$ and $\log \Sigma_5 < 0$, and $P_\mathrm{assoc} < 0.1$ and $\log \Sigma_5 > 0$ (grey circles, dots and stars, respectively).
}
\label{fig:fracquiestcent}
\end{figure}
In addition to colours, sSFR and SFR are also used to identify galaxies that have shut down their star formation  \citep[e.g.][]{peng2010, bluck2019}. These alternative proxies for quenching allows the selection of galaxies outside the SFMS, independently of their colours or morphology.  
 Here, we follow the criterion in \citet{peng2010}, which considers that a galaxy is quenched  when sSFR~$\leq 0.1$~Gyr$^{-1}$. This results in an average fraction of quiescent galaxies ($f_\mathrm{Q}$) of $28$\% in groups and $6$\% in the field. 
 These average values are similar, although slightly higher than the average fraction of red ($f_\mathrm{R}$) galaxies in groups ($23$\%) and in the field ($5$\%). 
 The difference between the quenched and red fractions can be accounted for by the small fraction of blue galaxies with sSFR~$\leq 0.1$~Gyr$^{-1}$. 
 
Both, $f_\mathrm{Q}$ and $f_\mathrm{R}$ change with \logM{}. Both show similar behaviour with \logM{}; although $f_\mathrm{Q}$ at the highest mass bins is higher than $f_\mathrm{R}$.This is mainly due to a larger fraction of quenched galaxies in groups that have still blue colours. However, $f_\mathrm{Q}$ and $f_\mathrm{R}$ for massive galaxies in the field is similar. This difference can be explained if there is a large fraction of post-starburst galaxies in dense environment with respect to the field, as found previously in clusters \citep{poggianti2009_postSB}. Post-starburst galaxies have shut down recently and rapidly their star formation, but they have still an intermediate age population dominating the optical colours.

To differentiate between the dependence of $f_\mathrm{Q}$ with \logM{} and with environment, Fig.~\ref{fig:fracquiestcent} shows $f_\mathrm{Q}$ vs \logM{} (in $0.4$~dex mass bins) for galaxies in groups and in the field. Although $f_\mathrm{Q}$ increases with \logM, for galaxies more massive than $10^{10}$~$M_\odot$, the increase is significantly higher for galaxies in groups than in the field. In groups $f_\mathrm{Q} \sim4$\% / 50\% / 80\%  for $\logMt\sim$10/$\sim$11/$\sim$11.5; while in the field $f_\mathrm{Q} \sim2$\% / 20\% / 55\%.


To check the dependence of $f_\mathrm{Q}$ with the criteria for group membership, we calculate $f_\mathrm{Q}$ after changing the threshold value of $P_\mathrm{assoc}$. The results indicate that $f_\mathrm{Q}$ in groups varies only a little when $P_\mathrm{} >$ 0.7 is changed to $P_\mathrm{assoc} > 0.8$ or $P_\mathrm{assoc} > 0.6$. In fact, $f_\mathrm{Q}$ goes from $79$~\% to $71$~\% for galaxies with $\logMt = 11.4$, which are the galaxies for which the difference in $f_\mathrm{Q}$ is higher. However, the variation of $f_\mathrm{Q}$ with the criteria to select galaxies in the field is more significant (Fig.~\ref{fig:fracquiestcent}). Here, we compare the results obtained with the following   criteria: (i) $P_\mathrm{assoc}< 0.1$; (ii) $P_\mathrm{assoc}< 0.1$ and $\log \Sigma_5 < 0$; (iii) $P_\mathrm{assoc} < 0.1$ and $\log \Sigma_5 > 0$. Although $f_\mathrm{Q}$ for low mass galaxies is independent of the field definition, $f_\mathrm{Q}$ changes significantly for the galaxies more massive than $\logMt > 11$. Thus, $f_\mathrm{Q}$ varies from $25$~\% to $53$~\% in the field at $\logMt = 11.4$. However, in each bin of galaxy stellar mass, the upper limit $f_\mathrm{Q}$ in the field sample is still significantly below than $f_\mathrm{Q}$ in groups. Thus, independently of the galaxy field selection, and the galaxy group population, $f_\mathrm{Q}$ is always higher in groups.

\subsection{Fraction excess of quenched galaxies}

\begin{figure}
\centering
\includegraphics[width=0.49\textwidth]{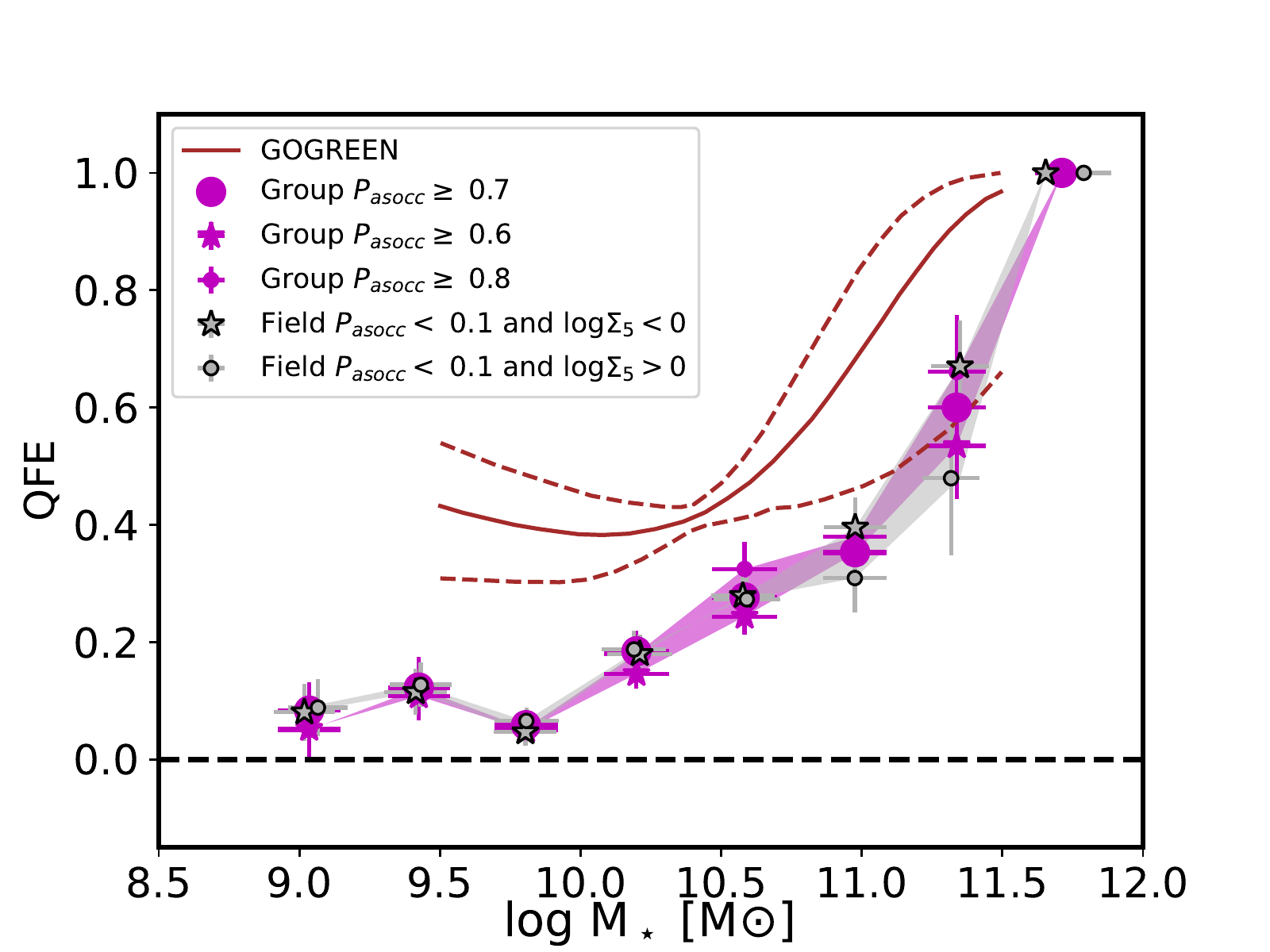}
\caption{ The quenched fraction excess in different bins of the stellar mass. Different symbols are the QFE measurements derived by the different assumptions to select galaxies members of groups, or galaxies in the field as in Fig. \ref{fig:fracquiestcent}. The brown curve is taken from \citet{mcnab2021}, the QFE inferred from the Schechter function fits to the GOGREEN data. The dashed lines represent their 68~\% confidence interval.
}
\label{fig:QFE}
\end{figure}

The most important characteristic of the galaxy populations in groups with respect to less dense environments is the large fraction of red galaxies. This characteristic is related to a larger fraction of galaxies in groups that have quenched their star formation  with respect to galaxies in the field. This difference can be measured by the quenched fraction excess (QFE). Here we adopt the definition given by \citet{mcnab2021} in their Eq.~(4):

\begin{equation}
\label{eq:QFE}
QFE =  (f^\mathrm{F}_\mathrm{SF} - f^\mathrm{G}_\mathrm{SF}) / f^\mathrm{F}_\mathrm{SF}\ ,
\end{equation}
where $f^\mathrm{F}_\mathrm{SF}$ and $f^\mathrm{G}_\mathrm{SF}$ are the fractions of star forming galaxies in the field and in groups, respectively. We note that a larger fraction of quenched galaxies is equivalent to a lack of star forming galaxies in groups with respect to low density environments.  QFE represents the fraction of field star forming galaxies that need to be quenched at the epoch of observation to be in the same abundance of that population in groups at the same time. This measures the quenching that can be attributed to the group environments. 
This definition is equivalent to the environmental quenching efficiency defined by \citet{peng2010, wetzel2015, nantais2017, vanderburg2018}, the transition fraction by \citet{vandenbosch2008}, and the conversion fraction  by \citet{balogh2016, fossati2017}. 

Here, we study the behaviour of QFE with \M{} (Figure~\ref{fig:QFE}). QFE increases with \M{} for galaxies above $10^{10}$~$M_\odot$, it is smaller than $\sim 10$~\% for galaxies less massive than $10^{10}$~$M_\odot$ and it is negligible below $10^{9}$~$M_\odot$. This behaviour is independent of the definition of field and the value of $P_\mathrm{assoc}$ to select galaxy members in the groups. For example, for galaxies of $10^{10}$ to $10^{11}$~$M_\odot$, QFE changes by $<5$~\%, when galaxies members of the groups are selected with $P_\mathrm{assoc} > 0.6$ or greater than $0.8$ instead of $0.7$. A similar effect is produced when only galaxies with $\Sigma_5 > 1$~Mpc$^{-2}$ or below this density are included in the sample of field galaxies, as well as \Pa{}$<0.1$. A larger effect is found for mass bins higher than $10^{11}$~$M_\odot$. In this range, QFE increases from $0.48$ to $0.67$ by changing the field definition.

The behaviour of QFE with \M{} found in our work  is similar to the one of QFE derived for high density environments \citep{balogh2016, vanderburg2020, mcnab2021}. However, we find relevant differences with respect to cluster environments \citep{mcnab2021, vanderburg2020}. For example, in the GOGREEN survey \citep{balogh2017}, that measures the rate of environmentally driven star formation quenching in clusters at $z\sim 1$, QFE is constant at $\sim 0.4$ for $\logMt < 10.5$ \citep{mcnab2021, vanderburg2020}; while we find that QFE in group environments is significantly smaller ($\leq 0.1$) for galaxies with $10^{10}$~$M_\odot$. For more massive galaxies in clusters, QFE increases up to $\sim 1$ for galaxies of $\logMt = 11.5$; while in AEGIS groups, QFE increases with the stellar mass but only up to $0.6$ for $\logMt = 11.5$, and rises to $0.4$ for $\logMt = 11$. 
It is interesting to note that QFE $\sim$0.4 is the value derived by \citet{peng2012} for low redshift satellite galaxies; and it is the value that we derive for galaxies of $10^{11}$~$M_\odot$.


\subsection{Fraction excess of transition galaxies}
\label{sec:transition}

It is interesting to identify galaxies that are in a transition phase between the blue cloud and the red sequence because this population provides clues for the rate of environmental quenching. Post-starburst galaxies, blue quiescent galaxies, red spirals, and green valley galaxies \citep{poggianti1999, tojeiro13, schawinski2014, lopes2016} form part of this transition galaxy populations.  

First, we have identified blue quiescent galaxies as those galaxies classified as blue due to their colour in the \urint{}--\logM{} plane and with $sSFR < 0.1$~Gyr$^{-1}$. They are below to the SFMS (see Fig.~\ref{fig:SFMS}). 
They are expected to be galaxies in a transition phase between the blue cloud and the red sequence. In fact, in the colour \urint{}--\logM{} diagram they are located in the area just below the red sequence (see Fig.~\ref{fig:Masscolour}), where the so-called green valley is \citep{schawinski2014}. 
The fraction of these galaxies in \mjp{} is small ($2.3$~\%), but is significantly larger in the group environments ($6.7$~\%) than in the field ($1.8$~\%). The fraction excess of these galaxies, defined as the difference between the fraction of transition galaxies in group with respect to the field, $f_\mathrm{i}^\mathrm{G}$-$f_\mathrm{i}^\mathrm{F}$, is not constant with \M{}. Although uncertainties are large due to the small number of transition galaxies in each bin of galaxy stellar mass, we found that the differential fraction ($\Delta f_\mathrm{i}$) is equal to zero for galaxies with $\logMt \geq 11$, but increases  from $5$ to $9$\%  for galaxies with masses between $10^{10.5}$  and $10^{9}$~$M_\odot$. These results are consistent with those obtained by \citet{mcnab2021} for the galaxy members of  clusters from the GOGREEN survey, where $f_\mathrm{i}^\mathrm{G}$-$f_\mathrm{i}^\mathrm{F}$ for the blue quiescent galaxy population increases to lower galaxy masses up to $\sim 10$~\%. 

Instead of colours and sSFR, some studies use the relative position of the galaxy with respect to the star forming main sequence to identify green valley objects. Here we use the \citet{bluck2020} definition, where green-transition galaxies are those that for a given mass bin have a variation with respect to the SFMS, $\Delta$SFR ($\log \mathrm{SFR}$ -- $\log \mathrm{SFR}_\mathrm{MS}$), between $-0.5$ and $-1$~dex. Using this definition and the SFMS law for star forming galaxies in groups and in the field derived in Sect.~\ref{sec:Mainsequence}, we identify this transition galaxy population. The fraction of this transition population and its behaviour with \logM{} is similar to the results derived for the blue quiescent population. Further, both criteria to identify the transition galaxy population provide a value of  $f_\mathrm{i}^\mathrm{G} - f_\mathrm{i}^\mathrm{F}$ that is compatible with the results derived in the GOGREEN survey \citep{mcnab2021}.


\begin{figure}
\centering
\includegraphics[width=0.49\textwidth]{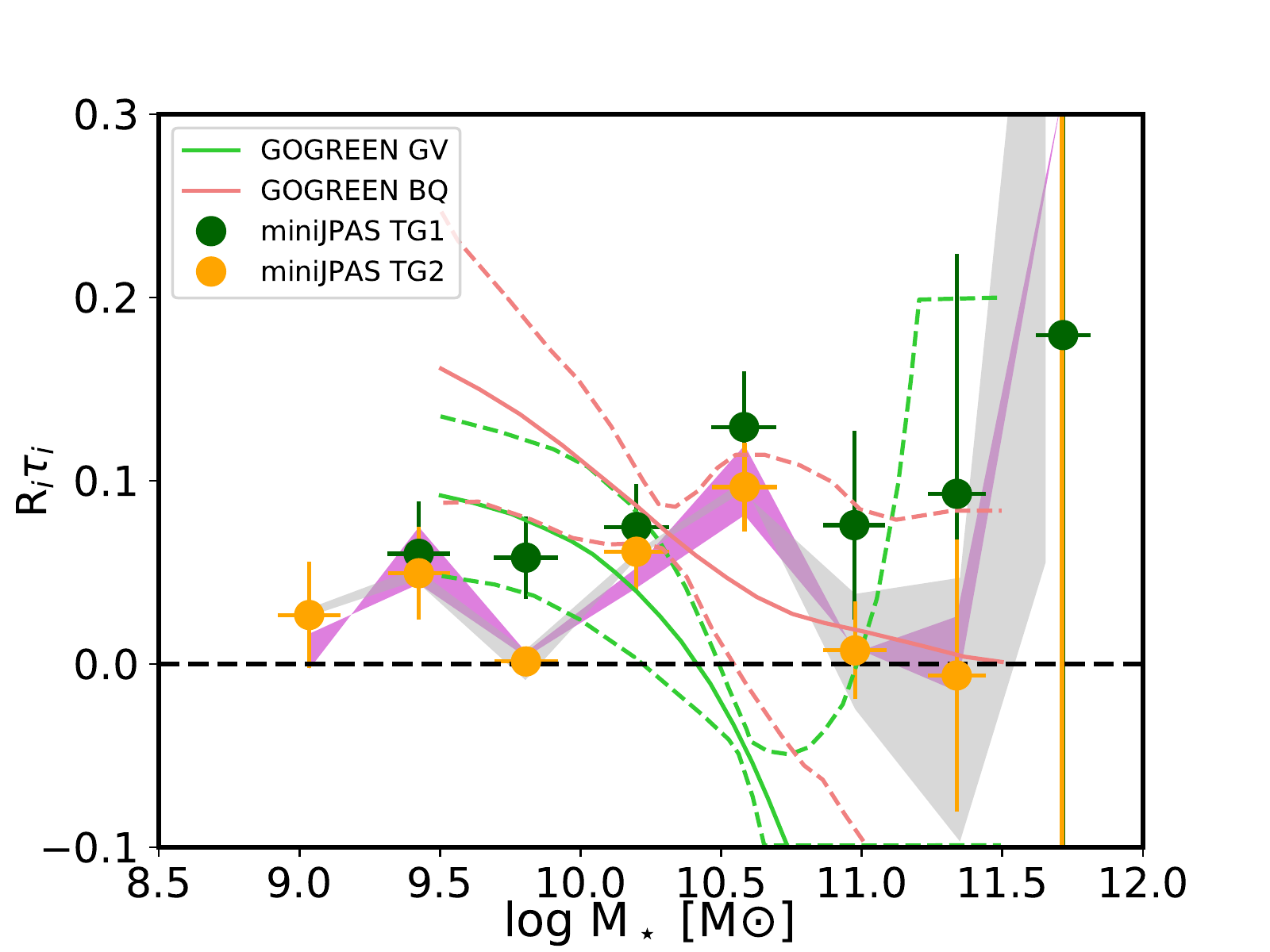}
\caption{$R_\mathrm{i} \tau_\mathrm{i}$ as a function of galaxy stellar mass, calculated from Eq.~(\ref{eq:rate}). Orange points represent the values obtained using the colour--sSFR definition for the transition galaxies (TG2). Green points represent the values obtained using the $\Delta$ SFR definition for transition galaxies (TG1). The magenta and grey shades correspond to the different assumptions on \Pa{} for the group members and field galaxies, respectively. Light green and coral lines are the results from \citet{mcnab2021} for the green and blue quiescent galaxy populations in GOGREEN. Dashed lines (same colours) represent the $68$\% confidence limits of their fit.
}
\label{fig:TransTime}
\end{figure}


The excess in the abundance of transition galaxies is related to the rate of the environmental quenching  and the time spent in that phase. If $R_\mathrm{i}$ is the fraction of field star forming galaxies that are quenched per unit of time as they are falling into the group, and $\tau_\mathrm{i}$ the time spent in the transition phase, the relative abundance excess of transition galaxies is: 
\begin{equation}
\label{eq:rate}
R_\mathrm{i} \tau_\mathrm{i} =  (f^\mathrm{G}_\mathrm{i} - f^\mathrm{F}_\mathrm{i}) / f^\mathrm{F}_\mathrm{SF}\ .
\end{equation}

As pointed out by \citet{mcnab2021},  this equation requires several assumptions: (i) the mass accretion rate is constant with time; (ii) the abundance excess of transition galaxies is produced only by quenching; (iii) the number of transition galaxies due to non-environmental reasons is proportional to the total galaxy populations.

Figure~\ref{fig:TransTime} shows $R_\mathrm{i} \tau_\mathrm{i}$ as a function of the stellar mass calculated with Eq.~(\ref{eq:rate}) assuming different definitions of the galaxies in groups and in the field. $P_\mathrm{assoc}$ changes from 0.6 to 0.8, and galaxies with $P_\mathrm{assoc} <$ 0.1 and $\log \Sigma_5$ $>0$ or $<$0 are selected to belong to field. As explained above, we take different assumptions for selecting transition galaxies based on the sSFR and \urint{} colour (magenta and grey points), or by $\Delta$SFR with respect to the MSSF (green points). It is difficult to estimate $R_\mathrm{i} \tau_\mathrm{i}$ for galaxies with masses larger than $10^{11}$~$M_\odot$. The value of $R_\mathrm{i} \tau_\mathrm{i}$ shows differences whether we select the transition galaxies according to the green valley or the blue quiescent galaxies.
For mass bins lower than $10^{11}$~$M_\odot$, $R_\mathrm{i} \tau_\mathrm{i}$ is not very dependent of the field or galaxy group members selection, neither of the transition galaxy population definition. There is an increase of $R_\mathrm{i} \tau_\mathrm{i}$ with decreasing mass, with a maximum of $R_\mathrm{i} \tau_\mathrm{i} \sim 0.1$  at $10^{10.5}$~$M_\odot$, and a plateau of $0.05$ for lower masses. 
The results are compatible with \citet{mcnab2021} for the GOGREEN survey. However, their results show an increase of $R_\mathrm{i} \tau_\mathrm{i}$ with decreasing mass, and the mass threshold for $R_\mathrm{i} \tau_\mathrm{i} >$ 0 is  $10^{11}$~$M_\odot$ for the blue quiescent galaxies and $10^{10.5}$~$M_\odot$ for the green valley populations.  It is worth pointing out that the $68$~\% confidence limits in $R_\mathrm{i} \tau_\mathrm{i}$ for the GOGREEN survey are compatible with the maximum $R_\mathrm{i} \tau_\mathrm{i}$ at $10^{10.5}$~$M_\odot$ in \mjp, as well as the lack of constraint in the most massive galaxies of the sample. 

\subsection{Transition time-scales}
\label{sec:time-scales}

To infer the rate of environmental quenching $R_\mathrm{i}$ we need to know the time that the transition galaxies spend in this phase, $\tau_\mathrm{i}$. Some works have suggested that $\tau_\mathrm{i} \sim 0.5$~Gyr, in particular for post-starburst and blue quiescent galaxies \citep{belli2019, muzzin2014}. In other works, $\tau_\mathrm{i}$ is equivalent to the time that the galaxy spends fading its star formation ($t_\mathrm{fade}$). In particular,  \citet{balogh2016} have estimated $t_\mathrm{fade}$ = $0.5 \pm 0.2$~Gyr for clusters and $0.9 \pm 0.3$~Gyr for groups from the surveys at $0.8 < z < 1.2$.

Here, we calculate $\tau_\mathrm{i}$ as the fading time-scale ($t_\mathrm{fade}$) that is related to  the relative abundance of the transition galaxies with respect to the star-forming galaxy population \citep{balogh2016}. Thus, 
\begin{equation}
\label{eq:trans}
t_\mathrm{fade}/t_\mathrm{SF+trans} = \tau_\mathrm{i}/t_\mathrm{q} = \tau_\mathrm{i}/T =
f^\mathrm{G}_\mathrm{i}  / (f^\mathrm{G}_\mathrm{SF} + f^\mathrm{G}_\mathrm{i})\ ,
\end{equation}
where $t_\mathrm{SF+trans}$ is the time during which all the presently star-forming and transition satellite galaxies fall into the cluster; it can be approximated to the lifetime ($T$) of the cluster at a given epoch, which is also equivalent to the total quenching-time scale ($t_\mathrm{q}$). 

\citet{balogh2016}, using the Millennium simulations \citep{springel2005}, and assuming that QFE evolves similarly to the fraction of halo mass assembled, find that $t_\mathrm{q}$ evolves like the dynamical time as $A \times (1+z)^{-3/2}$. $A$ is the lookback time when the halo started to assemble satellites, and it has a small dependence with the halo mass. Assuming that our AMICO groups have a halo mass of several times $10^{13}$~$M_\odot$, we derive $T = t_\mathrm{q} \sim 6.5$~Gyr for the mean redshift ($z = 0.39$) of our transition galaxies; and $\tau_i$ = $t_\mathrm{fade} \sim 0.8$~Gyr and $1.5$~Gyr if colour and sSFR (TG1) or $\Delta SFR$ (TG2) is used to select the transition galaxy population in \mjp. These values are in agreement with the average $t_\mathrm{fade}$ derived by \citet{balogh2016} for groups in a similar range of redshift.

\subsection{The evolution of group galaxy quenching}
\label{sec:evolutionratequenching}

\begin{figure}
\centering
\includegraphics[width=0.49\textwidth]{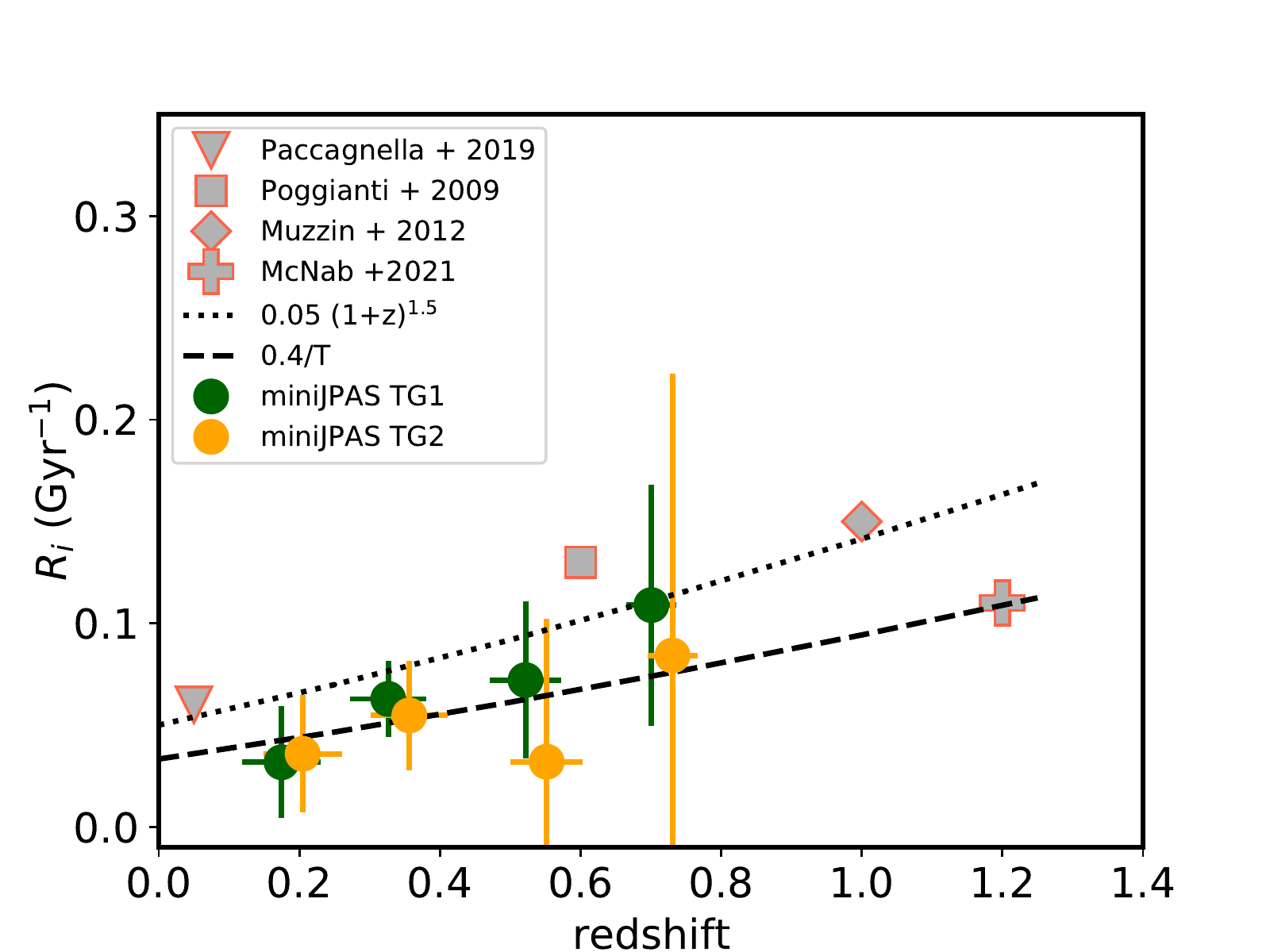}
\caption{Rate of group galaxy quenching in four redshift intervals for \mjp{}. The rate is calculated using two proxies for selecting  transition galaxies (orange and green points). Results from Fig.~15 in \citet{mcnab2021} are shown (grey-red points). Dashed line is $\mathrm{QFE}/\mathrm{T}$ for $\mathrm{QFE} = 0.4$.  Dotted line represents the evolution of the inverse of the dynamical time, $(1+z)^{3/2}$ scaled to $R=0.05$~Gyr$^{-1}$ at $z=0$  (see the text for further explanation).
}
\label{fig:evolutionrate}
\end{figure}

To explore the evolution of the group galaxy quenching rate, we divide the group sample in four redshift bins from $0.05$ to $0.85$ and width of $0.2$. First, we derive $T$ ($t_\mathrm{q}$) for the mean redshift within each bin, and $\tau_i$ ($t_\mathrm{fade}$) by using Eq.~(\ref{eq:trans}) for each redshift interval. Then, the evolution of the rate of the group galaxy quenching is derived from Eq.~(\ref{eq:rate}) and $\tau_i$ in the four redshift bins (Fig.~\ref{fig:evolutionrate}). Further, we calculate $T$ ($t_\mathrm{q}$), $\tau_i$ ($t_\mathrm{fade}$) and $R_i$ for the two proxies used to define transition galaxies: blue colour and sSFR $< 0.1$~Gyr$^{-1}$ (blue quiescent galaxies, magenta points) or $-1 < \Delta \mathrm{SFR} < -0.5$~dex (green valley, green points). 
It is remarkable that both proxies provide similar results for the group galaxy quenching rate ($R$) within the uncertainties, that are compatible with a modest but significant evolution in $R$ from $z\sim 0.8$ to $0.2$. This evolution is compatible with the expected evolution by a constant $\mathrm{QFE} = 0.4$; $R = \mathrm{QFE}/\mathrm{T}$ with $T$ being the life time of a cluster formed at $z=3$ at each given epoch (dashed line in Fig.~\ref{fig:evolutionrate}). This line also connects with the quenching rate $R$ derived for GOGREEN clusters at $z =1.2$ (grey-yellow cross) from \citet{mcnab2021}. Other results from \citet{paccagnella2019}, \citet{poggianti2009_postSB}, and \citet{muzzin2012} as adapted from \citet{mcnab2021} are shown in Fig.~\ref{fig:evolutionrate}.
The evolution of the inverse of the dynamical time, $(1+z)^{3/2}$,  scaled to $R= 0.05$~Gyr$^{-1}$ at $z= 0$ is also plotted.
Although our uncertainties are large, in particular at the two highest redshift bins, $R$ is below this evolutionary line. Thus, we conclude that the rate of group quenching shows a modest evolution that  
is compatible with a simple model in which QFE is constant and equal to 0.4 at $z\sim 0$. On the other hand, QFE is equal to 0.4 at $0.05 < z < 0.25$ in the \mjp{} groups. 

\subsection{The efficiency of group galaxy quenching}
\label{sec:efficiency}

\begin{figure}
\centering
\includegraphics[width=0.49\textwidth]{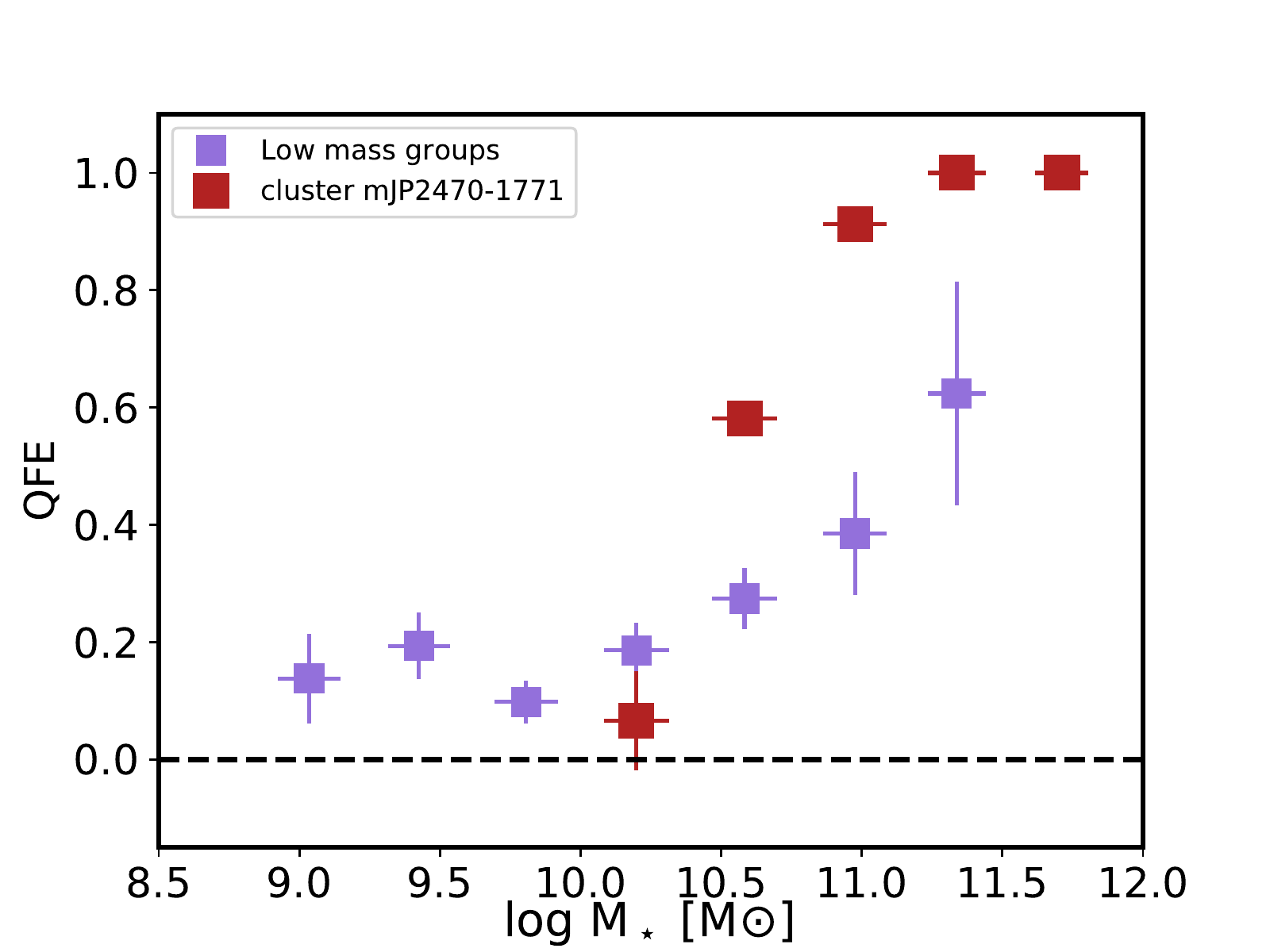}
\caption{The quenched fraction excess derived by using the lower massive groups ($M_{group} <$ 5$\times$10$^{11}$ M$\odot$) and for the more massive density structure in the \mjp{} survey, the cluster mJP2470-1771.
}
\label{fig:efficiency}
\end{figure}

Our results suggest that group environments are less efficient to quench galaxies that clusters, since QFE is about a factor $\sim$2 lower in the AEGIS groups than in clusters, at least for $M_\star <$10$^{11}$ M$\odot$ (see Fig. \ref{fig:QFE}). This is expected from the IllustrisTNG simulations that show that the quenched fraction increases with the virial mass, M$_{200}$ \citep{donnari2021a, donnari2021b}. They found that the fraction of  galaxies, with M$_\star \sim$ 10$^{10}$ M$\odot$, that are quenched at $z=0$ increases from $f_Q$ = 0.2 to 0.7 if they belong to halos of M$_{200}\sim$10$^{13}$ M$\odot$ to 10$^{14}$ M$\odot$. In contrast, $f_Q$ $\sim$0.9, independently of  M$_\star$, in very massive halos ($>$10$^{14}$ M$\odot$)  \citep{donnari2021a}.

To compare the efficiency of groups with respect to clusters, Fig. \ref{fig:efficiency} compares the QFE derived from the smaller groups in our sample ($M_{group} <$ 5$\times$10$^{11}$ M$\odot$) with respect to the most massive structure found by AMICO in \mjp{}, the cluster mJP2470-1771 \citep{bonoli21}. A very detailed analysis of the spatial distribution of galaxy populations (quenched, star-forming galaxies and AGNs) are presented in Rodr\'iguez-Mart\'in et al. (2022, subm). Its halo mass is $M_{200}=3.3\times 10^{14} ~ \mathrm{M}_\odot$. From our own analysis presented here, we estimate that this cluster is $\sim$ 10 times more massive than the most massive group considered in Fig. \ref{fig:efficiency}. We find that QFE is about a factor two higher in this cluster with respect to AMICO small groups, and in agreement what it is expected from the IllustrisTNG simulations. 

From this limited analysis, we can conclude that small groups are certainly less efficient in  quenching galaxies with $M_\star >$10$^{10}$ M$_\odot$ than clusters.
Small groups seem to be also efficient to quench galaxies with $M_\star <$10$^{10}$ M$_\odot$.
These results can be understood because galaxies can be quenched after being accreted into the cluster-host. A deeper potential well associated to the cluster stimulates galaxy interactions and mergers and tidal stripping; higher galaxy number density favor harassment; and higher ambient gas density advantages ram-pressure stripping processes. Further, galaxies can be also quenched while they are members of pre-processing group hosts that are also accreted to the cluster.

\section{Summary and conclusions}
\label{sec:conclusions}

The goal of this paper is to illustrate the power of \jp{} to investigate the role that environment plays in galaxy evolution. In particular, the role of group  environment in quenching the star formation in galaxies, and its evolution since $z\sim 1$. First, 
we analysed the stellar population properties of galaxies with $M_\star \geq$ 10$^9$ M$\odot$ that belong to a sample of 80 groups with M$_{200}$ down to 10$^{13}$ M$\odot$ detected by AMICO in \mjp. Galaxy members of the groups were selected based on the association probability from AMICO, $P_{\mathrm{assoc}} \geq 0.7$.  A sub-sample of galaxies of \mjp{} to be representative of a low number density field ($P_\mathrm{assoc}< 0.1$) was also selected. Using the parametric SED-fitting code \baysea{}, we derived the star formation history and the stellar population properties of the galaxies that belong to groups or to the field. In particular, 
we used the extinction corrected rest-frame colour \urint{}, \M{}, and sSFR, for the identification of galaxies that have shut down their star formation. We measured the abundance of
the red and blue galaxy populations, and the transition galaxies that are in a phase between the blue cloud and the red sequence as a function of the galaxy stellar mass in groups and in the field. 
Uncertainties associated to the criteria for the selection of galaxies as a function of the environment density were also studied. The main conclusions are: 

\begin{itemize}
    \item {{\it Stellar population properties:} Galaxies in groups are slightly older, redder, more metal-rich, with lower sSFR, and smaller $\tau/t_0$ values than galaxies in the field. However, the red galaxy population in groups has similar properties to that of the equivalent galaxy population in the field. Further, the evolution of the properties since $z\sim$1 is similar for the group and field galaxy populations. }
    
    \item {{\it BGG:} The central (brighest) galaxy (BGG) is the most massive galaxy of the group. These galaxies are among the oldest, reddest, and more metal-rich objects of the red galaxy population. They have on average sSFR below 0.1 Gyr$^{-1}$, and $\tau/t_0$ = 0.17$\pm$0.34, indicating that the star formation has already shut down, as most of the red-galaxy populations in groups and in the field.  }
    \item {{\it Abundance of red galaxies:} The fraction of red galaxies in \mjp{} increases with the galaxy stellar mass; although this fraction is always higher in groups than in the field for galaxies more massive than $10^{10}$~$M_\odot$. The fraction of blue galaxies decreases as the galaxy mass increases, and evolves as expected with redshift, being lower at $z\sim$0.1 than at $z\sim 0.8$.}
    \item {{\it Fraction of quenched galaxies:} $28$~\% of the group galaxy population has $sSFR \leq 0.1$~Gyr$^{-1}$. This fraction is almost independent of the threshold value of $P_\mathrm{assoc}$ used  to define the galaxy group members. In contrast, the fraction of quenched galaxies in the field is only $5$~\%, and it shows a dependence with the field galaxy population selected for the  galaxies more massive than $10^{11}$~ $M_\odot$. }
    \item {{\it Quenched fraction excess:} The QFE shows a strong dependence with galaxy stellar mass above $10^{10}$~$M_\odot$, increasing from a few percent for $M_\star < 10^{10}$~$M_\odot$, to $40$~\% for $10^{11}$~$M_\odot$, and $60$~\% for $10^{11.5}$~$M_\odot$. 
    }
    \item {{\it Transition galaxies:} Blue quiescent and green valley galaxies are identified as transition galaxy populations based on their colours, sSFR, or their SFR offset with respect to the main sequence of the star-forming galaxies in groups and in the field. The fraction of transition galaxies is higher in group environment than in the field.}
    \item {{\it $R_\mathrm{i} \tau_\mathrm{i}$:}
    The abundance excess of transition galaxies relative to the star-forming galaxy population ($R_\mathrm{i} \tau_\mathrm{i}$) shows a slight dependence with galaxy stellar mass, found between $10$ to $5$~\% for galaxies less massive than $10^{11}$~$M_\odot$.    
    }
    \item {{\it Transition time scale:} It is defined as the fading time scale ($t_\mathrm{fade} = \tau_\mathrm{i}$ ). It depends on the abundance of transition galaxies relative to the star-forming galaxies and the time since the transition and star-forming galaxies were accreted  to form the group. We obtain a mean value for $t_\mathrm{fade} \sim 0.8$~Gyr for blue quiescent galaxies and $\sim 1.5$~Gyr for the green valley galaxies.}
    \item {{\it The evolution of galaxy quenching rate:} We find  that $R_\mathrm{i}$ shows a modest but significant evolution since $z \sim 0.8$. This evolution is compatible with the expected evolution by a constant $QFE=0.4$, and it connects with the expected evolution for clusters at $z = 1$.
    }
    
\end{itemize}

These results show the potential of \jp{} to constrain the role that groups and clusters play on galaxy evolution. This potential resides in the accurate  \photoz{} estimations \citep[]{hernan-caballero21} that allow us to identify groups and clusters up to $z < 1$ to produce unbiased and complete mass-sensitive catalogues. Further, the \js{} allow us to retrieve stellar population properties for the blue and red galaxy populations with a precision similar to future spectroscopic surveys with similar S/N as we have proven in \citet{gonzalezdelgado21}. Volume complete  samples can be studied above of $\log (M_\star/\mathrm{M}_\odot) \sim 8.9$, $9.5$, and $9.9$ at $z=0.3$, $z=0.5$ and $z=0.7$, respectively.

Based on the whole sample analysed here, we expect that \jp{} will detect more than 90 millions of galaxies with $r < 22.75$~AB for which the stellar population properties will be derived. This is more than a factor 20 over the whole SDSS survey (approximately 4.3 million spectra of galaxies over the same area), and nearly a factor 3 larger that the number of galaxies that will be 
observed by DESI (33 million galaxies over $14000$~deg$^2$). In addition, more than 0.6 millions of groups and clusters will be detected (Maturi et al. in prep.), and it will be possible to disentangle the quenching due to the halo mass and the environment in a wide range of galaxy stellar mass, groups, clusters properties and the evolution since $z\sim 1$.


\begin{acknowledgements} 

R.G.D., L.A.D.G., R.G.B., G.M.S., J.R.M., and E.P. acknowledge financial support from the State Agency for Research of the Spanish MCIU through the `Center of Excellence Severo Ochoa' award to the Instituto de Astrof\'isica de Andaluc\'ia (SEV-2017-0709), and to PID2019-109067-GB100.

I.M. acknowledges to PID2019-106027GB-C41.

S.B.  acknowledges to PGC2018-097585- B-C22, MINECO/FEDER, UE of the Spanish Ministerio de Economia, Industria y Competitividad. 

LSJ acknowledges the support from CNPq (304819/2017-4) and FAPESP (2019/10923-5).

Based on observations made with the JST/T250 telescope and the Pathfinder camera for the miniJPAS project at the Observatorio Astrof\'{\i}sico de Javalambre (OAJ), in Teruel, owned, managed, and operated by the Centro de Estudios de F\'{\i}sica del  Cosmos de Arag\'on (CEFCA). We acknowledge the OAJ Data Processing and Archiving Unit (UPAD) for reducing and calibrating the OAJ data used in this work.

Funding for OAJ, UPAD, and CEFCA has been provided by the Governments of Spain and Arag\'on through the Fondo de Inversiones de Teruel; the Arag\'on Government through the Research Groups E96, E103, and E16\_17R; the Spanish Ministry of Science, Innovation and Universities (MCIU/AEI/FEDER, UE) with grant PGC2018-097585-B-C21; the Spanish Ministry of Economy and Competitiveness (MINECO/FEDER, UE) under AYA2015-66211-C2-1-P, AYA2015-66211-C2-2, AYA2012-30789, and ICTS-2009-14; and European FEDER funding (FCDD10-4E-867, FCDD13-4E-2685).

JAFO acknowledges the financial support from the Spanish Ministry of Science and Innovation and the European Union -- NextGenerationEU through the Recovery and Resilience Facility project ICTS-MRR-2021-03-CEFCA.

\end{acknowledgements}




\bibliographystyle{aa}
\bibliography{AMICO_bibliography.bib}


\begin{appendix}

\end{appendix}

\end{document}